\documentclass[longauth]{aa}  
\usepackage{graphicx}
\usepackage{txfonts}
\usepackage[export]{adjustbox}
\usepackage{subfig}

\usepackage{longtable}
\usepackage{tabularx}
\usepackage{pdflscape}
\usepackage{multicol}
\usepackage{multirow}
\usepackage{hyperref}
 \hypersetup{
     colorlinks=true,
     linkcolor=blue,
     filecolor=magenta,
     urlcolor=cyan,
     citecolor=cyan,
 }
\usepackage{epsf}
\usepackage{rotating}

\usepackage{xcolor}

\usepackage{mathabx}

\usepackage[multiple]{footmisc}
\usepackage{siunitx}
\def\num#1{\numx#1}\def\numx#1e#2{{#1}\mathrm{e}{#2}}

\DeclareMathOperator{\e}{e}

\newcommand{\teff}{\mbox{$T_{\rm eff}$}}
\newcommand{\logg}{\mbox{$\log g$}}

\newcommand{\vsini}{\mbox{$v \sin i_*$}}

\newcommand{\kms}{\mbox{km\,s$^{-1}$}}
\newcommand{\ms}{\mbox{m\,s$^{-1}$}}

\newcommand{\MSun}{\mbox{$\mathrm{M}_\Sun$}}

\newcommand{\MEarth}{\mbox{$\mathrm{M}_\Earth$}}
\newcommand{\REarth}{\mbox{$\mathrm{R}_\Earth$}}

\providecommand{\abs}[1]{\lvert#1\rvert}

\newcommand{\tess}{TESS}

\newcommand{\rv}{RV}

\newcommand{\project}[1]{\textsf{#1}}

\newcommand{\emcee}{\project{emcee}}

\newcolumntype{P}[1]{>{\centering\arraybackslash}p{#1}}
\newcolumntype{L}[1]{>{\raggedleft\arraybackslash}p{#1}}

\newenvironment{symbolfootnotes}
  {\par\edef\savedfootnotenumber{\number\value{footnote}}
   
   \setcounter{footnote}{0}}
  {\par\setcounter{footnote}{\savedfootnotenumber}}
  
\usepackage{txfonts}
\begin{document}

   \title{HD~23472: A multi-planetary system with three super-Earths and two potential super-Mercuries \thanks{Based in part on Guaranteed Time Observations collected at the European Southern Observatory under ESO programme(s) 1102.C-0744, 1102.C-0958, and 1104.C-0350 by the ESPRESSO Consortium.} }
   \author{S.~C.~C.~Barros\inst{\ref{IA-Porto},\ref{UP}}\thanks{E-mail: susana.barros@astro.up.pt} 
  \and O. D. S. Demangeon \inst{\ref{IA-Porto},\ref{UP}}        
  \and Y. Alibert \inst{\ref{Bern}}
  \and A. Leleu \inst{\ref{Geneve-obs}}
   \and V. Adibekyan \inst{\ref{IA-Porto},\ref{UP}}
  \and C. Lovis \inst{\ref{Geneve-obs}}
   \and D. Bossini \inst{\ref{IA-Porto}}
   \and S. G. Sousa \inst{\ref{IA-Porto}}
   \and N. Hara \inst{\ref{Geneve-obs}}
   \and F. Bouchy \inst{\ref{Geneve-obs}}
   \and B. Lavie \inst{\ref{Geneve-obs}}
   \and J. Rodrigues \inst{\ref{IA-Porto},\ref{UP}}
 \and J. Gomes da Silva \inst{\ref{IA-Porto}}
 \and J. Lillo-Box \inst{\ref{CAB2}}
\and F. A. Pepe \inst{\ref{Geneve-obs}}
\and H. M. Tabernero \inst{\ref{CAB},\ref{IA-Porto}}
\and M. R. Zapatero Osorio \inst{\ref{CAB}}
\and A. Sozzetti \inst{\ref{INAF-Torino}}
\and A. Su\'arez Mascare\~no \inst{\ref{IAC},\ref{ULL}}
\and G. Micela \inst{\ref{INAF-Palermo}}
\and C. Allende Prieto \inst{\ref{IAC},\ref{ULL}}
\and S. Cristiani \inst{\ref{INAF-Trieste}}
\and M. Damasso \inst{\ref{INAF-Torino}}
\and P. Di Marcantonio \inst{\ref{INAF-Trieste}}
\and D. Ehrenreich \inst{\ref{Geneve-obs}}
\and J. Faria \inst{\ref{IA-Porto},\ref{UP}}
\and P. Figueira \inst{\ref{ESO-Chile},\ref{IA-Porto}}
\and J. I. Gonz\'alez Hern\'andez \inst{\ref{IAC},\ref{ULL}}
\and J. Jenkins \inst{\ref{AMES}}
\and G. Lo Curto \inst{\ref{ESO-Germany}}
\and C. J. A. P. Martins \inst{\ref{IA-Porto}, \ref{CAUP}}
\and G. Micela\inst{\ref{INAF-Palermo}}
\and N. J. Nunes \inst{\ref{IA-Lisboa}}
\and E. Pall\'e \inst{\ref{IAC},\ref{ULL}}
\and N. C. Santos \inst{\ref{IA-Porto},\ref{UP}}
\and R. Rebolo \inst{\ref{IAC},\ref{ULL},\ref{CIC}}
\and S. Seager  \inst{\ref{MIT},\ref{MIT-kavli},\ref{MIT-AA}}
\and  J. D. Twicken \inst{\ref{AMES}, \ref{SETI}}
\and S. Udry \inst{\ref{Geneve-obs}} 
\and R.~Vanderspek \inst{\ref{MIT-kavli}}
\and J. N.\ Winn \inst{\ref{Princeton}}
             }
 \institute{Instituto de Astrof\'isica e Ci\^encias do Espa\c{c}o, Universidade do Porto, CAUP, Rua das Estrelas, PT4150-762 Porto, Portugal \label{IA-Porto} 
             \and
            Departamento\,de\,Fisica\,e\,Astronomia,\,Faculdade\,de\,Ciencias,\,Universidade\,do\,Porto,\,Rua\,Campo\,Alegre,\,4169-007\,Porto,\,Portugal \label{UP}
\and Physics Institute, University of Bern, Sidlerstrasse 5, 3012 Bern, Switzerland \label{Bern}
 \and Instituto de Astrof\'isica e Ci\^encias do Espa\c{c}o, Faculdade de Ci\^encias da Universidade de Lisboa, Campo Grande, PT1749-016 Lisboa, Portugal
\label{IA-Lisboa}
\and Centro de Astrof\'isica  da Universidade do Porto, Rua das Estrelas, 4150-762 Porto, Portugal  \label{CAUP} 
\and D\'epartement d’astronomie de  l'Universit\'e de Gen\`eve, Chemin Pegasi, 51, 1290 Sauverny, Switzerland  \label{Geneve-obs}
\and Instituto de Astrof\'{\i}sica de Canarias (IAC), Calle V\'{\i}a L\'actea s/n, E-38205 La Laguna, Tenerife, Spain
\label{IAC}
\and Departamento de Astrof\'{\i}sica, Universidad de La Laguna (ULL), E-38206 La Laguna, Tenerife, Spain
\label{ULL}
\and Consejo Superior de Investigaciones Cient\'{\i}cas, Spain
\label{CIC}
\and Centro de Astrobiolog\'\i a (CSIC-INTA), Crta. Ajalvir km 4, E-28850 Torrej\'on de Ardoz, Madrid, Spain
\label{CAB}
\and Centro de Astrobiolog\'ia (CAB, CSIC-INTA), Depto. de Astrof\'isica, ESAC campus, 28692, Villanueva de la Ca\~nada (Madrid), Spain \label{CAB2}
\and INAF - Osservatorio Astronomico di Trieste, via G. B. Tiepolo 11, I-34143 Trieste, Italy
\label{INAF-Trieste}
\and INAF - Osservatorio Astrofisico di Torino, via Osservatorio 20, 10025 Pino Torinese, Italy
\label{INAF-Torino}
\and INAF - Osservatorio Astronomico di Palermo, Piazza del Parlamento 1, I-90134 Palermo, Italy
\label{INAF-Palermo}
\and European Southern Observatory, Alonso de C\'ordova 3107, Vitacura, Regi\'on Metropolitana, Chile
\label{ESO-Chile}
\and European Southern Observatory, Karl-Schwarzschild-Strasse 2, 85748, Garching b. M\"unchen, Germany
\label{ESO-Germany}
 \and  Department of Earth, Atmospheric, and Planetary Sciences, Massachusetts Institute of Technology, Cambridge, MA 02139, USA \label{MIT}               
 \and  Department of Physics and Kavli Institute for Astrophysics and Space Research, Massachusetts Institute of Technology, Cambridge, MA 02139, USA \label{MIT-kavli}         
 \and  Department of Aeronautics and Astronautics, Massachusetts Institute of Technology, Cambridge, MA 02139, USA\label{MIT-AA}     
\and Department of Astrophysical Sciences, Princeton University, Princeton, NJ 08544, USA  \label{Princeton}  
\and NASA Ames Research Center, Moffett Field, CA 94035, United States of America \label{AMES}  
\and  SETI Institute, Mountain View, CA  94043, USA  \label{SETI}  
       }
  
   \date{Received ??, ??; accepted ??} 
   \abstract
   {Comparing the properties of planets orbiting the same host star, and thus formed from the same accretion disc, helps in constraining theories of exoplanet formation and evolution. As a result, the scientific interest in multi-planetary systems is growing with the increasing number of detections of planetary companions. }
   { We report the characterisation of a multi-planetary system composed of five exoplanets orbiting the K-dwarf HD~23472 (TOI-174). }     
   {In addition to the two super-Earths that were previously confirmed, we confirm and characterise three Earth-size planets in the system using ESPRESSO radial velocity observations. The planets of this compact system have periods of $P_d \sim 3.98\,$, $P_e \sim 7.90\,$, $P_f \sim 12.16\,$,   $P_b \sim 17.67,\,$  and $P_c \sim 29.80\,$days and radii of  $R_d  \sim 0.75\,$ ,  $R_e  \sim 0.82,$, $R_f  \sim 1.13\,$, $R_b  \sim 2.01,\,$ and, $R_c  \sim 1.85\,$  \REarth. Because of its small size, its proximity to planet d's transit, and close resonance with planet d, planet e was   only recently found. }
   { The planetary masses were estimated to be $M_d =0.54\pm0.22$,  $M_e =0.76\pm0.30$,  $M_f =0.64_{-0.39}^{+0.46}$,  $M_b = 8.42_{-0.84}^{+0.83}$, and  $M_c = 3.37_{-0.87}^{+0.92}$ \MEarth. These planets are among the lightest planets, with masses measured using the radial velocity method, demonstrating the very high precision of the ESPRESSO spectrograph. We estimated the composition of the system's five planets and found that their gas and water mass fractions increase with stellar distance, suggesting that the system was shaped by irradiation. The high density of the two inner planets ($\rho_d = 7.5_{-3.1}^{+3.9}$ and $\rho_e = 7.5_{-3.0}^{+3.9}\, \mathrm{g.cm^{-3}}$) indicates that they are likely to be super-Mercuries. This is supported by the modelling of the internal structures of the planets, which also suggests that the three outermost planets have significant water or gas content.}
   { If the existence of two super-Mercuries in the system is confirmed, this system will be the only one known to feature two super-Mercuries, making it an excellent testing bed for theories of super-Mercuries formation. Furthermore, the system is close to a Laplace resonance, and further monitoring could shed light on how it was formed. Its uniqueness and location in the continuous viewing zone of the James Webb space telescope will make it a cornerstone of future in-depth characterisations.} 
   
 \keywords{planetary systems: fundamental parameters --planetary systems:composition  --techniques: photometric --methods:data analysis}

 \maketitle
%

\section{Introduction} 
\label{intro}
   
New high-resolution spectrographs, such as the Echelle Spectrograph for Rocky Exoplanet- and Stable Spectroscopic Observations (ESPRESSO) \citep{Pepe2021}, in conjunction with photometric space missions focusing on bright stars, such as K2 \citep{Borucki2010,Howell2014}, transiting exoplanet survey satellite (TESS) \citep{Ricker2015}, and CHaracterising ExOPlanet Satellite (CHEOPS)  \citep{Benz2021},  are pushing the limits of planet characterisation to planets of similar size and mass to Earth. This provides an unprecedented view of the interiors of small exoplanets \citep[e.g.][]{Silva2022, Demangeon2021,Toledo-Padron2020, Lillo-Box2020}. These studies are also pushing the boundaries of small planet characterisations to include cooler planets, thus allowing us to gain a better understanding of how high stellar irradiation shapes a planet's composition.

Multi-planetary systems are especially valuable because they share the same host star and were formed by the same accretion disc \citep[e.g.][]{Ormel2017,Grimm2018}. The properties of multi-planetary systems have been used to constrain planet formation and evolution models. For example, it has been demonstrated that the sizes of planets in a multi-planetary system are correlated and that there is regular spacing between planets \citep{Lissauer2011b, Ciardi2013, Weiss2018}, which is referred to as the  "peas in a pod" theory.  
According to a recent study, multi-planetary systems appear to be less similar in mass than in radius \citep{Otegi2022}. However, a larger sample of multi-planetary systems with well-characterised mass and radius is required to confirm this result and uncover additional correlations.

 High-energy irradiation received by short period planets causes evaporation of H/He-rich envelopes \citep[e.g.][]{Lammer2003,Yelle2004,Owen2017,Jin2018}. Atmospheric evaporation has been observed for the low-mass planet GJ~436b \citep{Kulow2014, Ehrenreich2015,Lavie2017}. As a result of irradiation close-in planets become denser and smaller \citep[e.g.][]{Lopez2012,Howe2015}. The evaporation theory predicted the existence of a gap in the radius distribution of planets \citep{Owen2013}. Such  a gap was later uncovered using the Kepler sample \citep{Fulton2017}. The California-Kepler Survey revealed that the distribution of planet sizes is bimodal and that there is a radius gap at a planetary radius of $\sim 1.6-1.8$  \REarth\ \citep{Fulton2017,VanEylen2018}.  Exoplanets with less than $2$  \REarth\ are thought to be dry naked cores, whereas exoplanets with more than $2$  \REarth\ are thought to have icy cores and possibly a gaseous atmosphere \citep{Venturini2020}.  Alternatively, the radius gap can also be explained by the theory of core-powered evaporation  \citep[e.g.][]{Ginzburg2018,Gupta2019}. Characterising multi-planetary systems with planets above and below the radius gap can shed light on the mechanism of mass loss that is responsible for these effects.

Some of these naked cores are extremely dense and likely contain an excess of iron, similar to Mercury. Since the discovery of the first super-Mercury, K2-229 b, \citep{Santerne2018} other exoplanets with an excess of iron have been uncovered: K2-38 b \citep{Toledo-Padron2020},  K2-106 b \citep{Guenther2017}, Kepler-107 c \citep{Bonomo2019}, Kepler-406 b \citep{Marcy2014}, and HD 137496 b \citep{Silva2022}. These planets, with the exception of Kepler-107 c, are all the inner planets of their exoplanetary system. They are all part of multi-planetary systems and have high effective temperatures ($> 1200$K). Several theories on planet formation and evolution are under discussion to explain their existence, including a giant impact \citep{Benz2007}, mantle evaporation \citep{Cameron1985}, photophoresis \citep{Wurm2013}, and the possibility of a compressed planetary core \citep{Mocquet2014}.  

We chose to characterise HD~23472 (TOI-174) because it serves as an ideal object for gaining further insight into atmosphere evaporation. When we began our observations, the system was reported to have two planets above as well as two planets below the radius gap in the EXOFOP-TESS database \citep{TESSexofop}. Furthermore, the planets span a range of stellar incident flux between 8 and 117 of that of the Earth. Hence, HD~23472 evidently stands as a golden target for mass characterisation with ESPRESSO. The new planet that we found below the radius gap makes the system even more interesting. Here, we present the characterisation of the five planets in the HD~23472 system, two of which are likely super-Mercuries. We present our new ESPRESSO radial observations of HD~23472 as well as previous RV and photometric observations in Section~\ref{observations}.  We describe our data analysis method in Section~\ref{analysis} and we present the derived system parameters in Section~\ref{modelfit}. In Section~\ref{dinamics}, we present our analysis of the system's dynamics, which allows us to better constrain the eccentricity of the planets, while in Section~\ref{composition} we present the analysis of the internal structure of the planets. Finally, our conclusions are summarised in Section~\ref{conclusion}. 

%
%
\section{Observations}
\label{observations}
\subsection{ESPRESSO observations}
\label{espressoobservations}

From 20 July 2019 to 9 April 2021, we collected 104 spectra of HD~23472 (K4V, V mag = 9.7) with ESPRESSO \citep{Pepe2021}, each with an exposure time of 900s. The ESPRESSO high resolution echelle spectrograph is mounted on the ESO Paranal Observatory's Very Large Telescope (VLT). ESPRESSO can collect light from any VLT unit telescope or all of them at the same time. The observations were obtained as part of the ESPRESSO Guaranteed Time Observations (programs 1102.C-0744, 1102.C-0958, and 1104.C-0350) which has already allowed to constrain the composition of several small planets \citep{Damasso2020, Toledo-Padron2020, Mortier2020,Sozzetti2021,Demangeon2021}.  
ESPRESSO is isolated in order to maintain constant pressure, temperature, and humidity. We used the single UT high resolution mode (HR11, fast-readout) for all observations which has a spectral resolution of R$=140\,000$ and covers wavelengths from 380 nm to 788nm. We obtained a S/N of 80 per resolution element at 650nm. All measurements were obtained with the Fabry Perrot in Fibre B for simultaneous calibration.This allows the correction of the instrumental drift with a precision better than $10$ cm/s \citep{Wildi2010}.

To extract the radial velocities (RVs), we used the version 2.21 of the ESPRESSO pipeline Data-reduction Software (DRS). The RVs are calculated using the DRS by cross-correlating the spectra with a stellar line mask \citep{Baranne1996}, which in our case was designed for K6 type stars. Then the DRS fits the cross correlating function (CCF) with an inverted Gaussian profile to obtain the centre of the Gaussian (RV measurement), the full width at half maximum (FWHM), and the amplitude (contrast of the CCF). The RVs' uncertainties are computed using the \citet{Bouchy2001} technique. Moreover, the DRS also computes other activity indicators: the BIS \citep{Queloz2001}, the depth of the $H_\alpha$ line \citep{Mascareno2015}, the depth of the Sodium doublet (NaD, \citealt{Diaz2007}),  and the S-index \citep{Lovis2011, Noyes1984}. The median uncertainty of the RV measurements is 0.38 m/s, and their peak-to-peak amplitude is 16.33 m/s.

\begin{figure*} 
\centering 
\includegraphics[width=2.0\columnwidth]{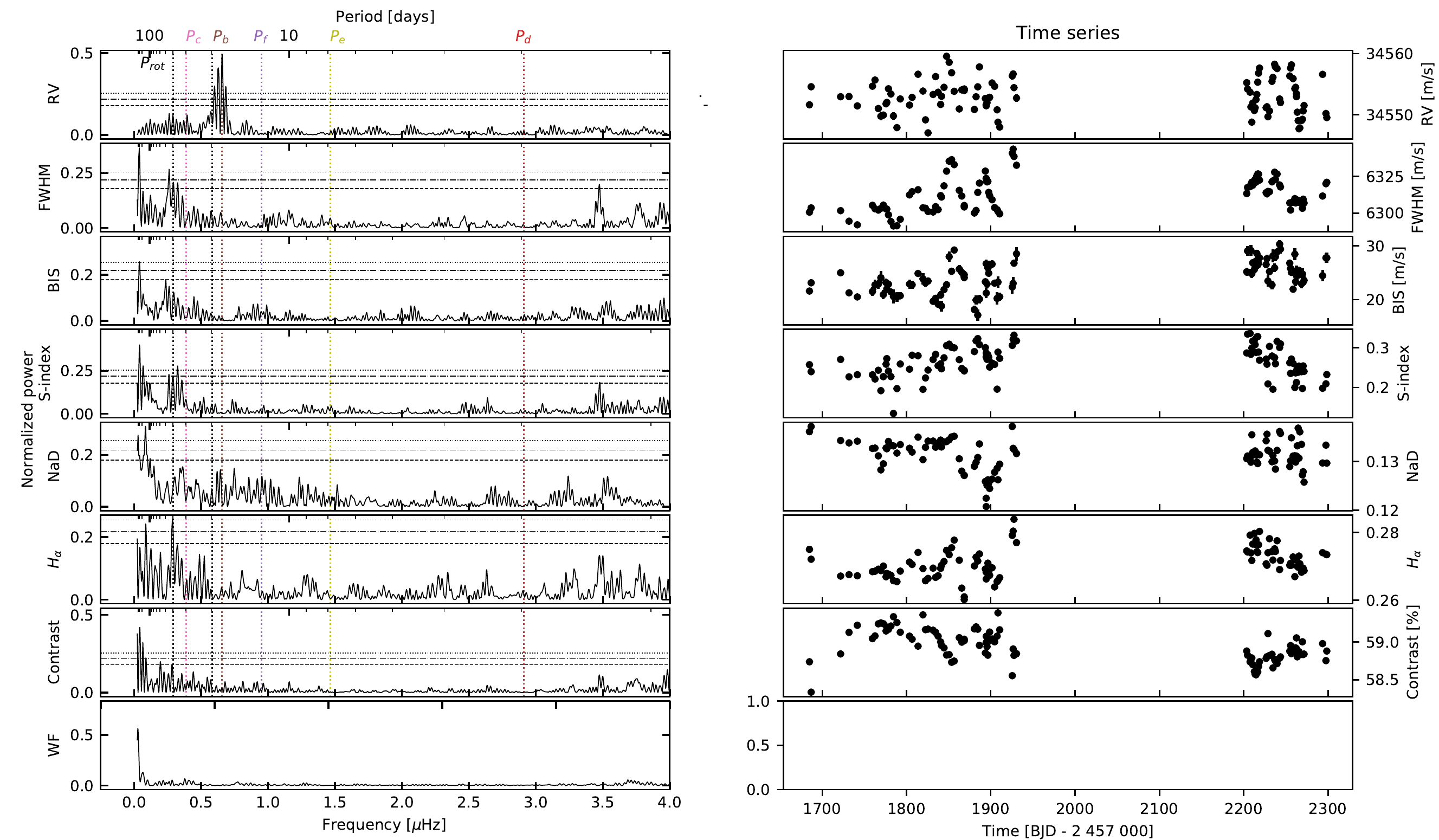}
\caption{Left panel: GLS of the RVs and indicators of the new ESPRESSO observations. The last row shows the window function. The coloured dotted vertical lines show the position of the known transiting planets, while the black dotted vertical lines show the position of the estimated rotation period of the star and its first harmonic. The horizontal lines indicate the 10\% (dashed line), 1\%(dot-dashed line), and 0.1\% (dotted line) FAP levels calculated following \citet{Zechmeister2009}. Right panel: Time series of the RV observations and the activity indicators.
\label{indicators}} 
\end{figure*}

The generalised Lomb Scargle Periodogram  (GLS, \citealt{Zechmeister2009}) of the RVs, activity indicators (FWHM, BIS, S-index, NaD,  $H_\alpha$), contrast, and the associated window function are shown in Figure~\ref{indicators}. We also show the time series of the RV and activity indicators. The two strongest peaks in the RVs correspond to the known planet b and c that have an orbital period of 17 days and 29.9 days, respectively. There is not a strong evidence for these periodicities in the indicators.  The third strongest peak in the RVs $\sim 40$ days, on the other hand, is most likely due to stellar activity, as it is also present in all of the indicators except the NaD.

\subsection{Previous RV observations}

Following the public release of transiting candidates in the EXOFOP-TESS database \citep{TESSexofop},  \citet{Trifonov2019} analysed 14 public HARPS RV measurements of HD~23472 to constrain the mass of the outermost planets ($P_b= 17.7$ days and $P_c=29.8$ days). The HARPS RVs allowed them to set an upper limit on the RV semi-amplitude of both planets: $K_b=5.33^{+0.67}_{-4.2}\,$\ms\  and $K_c=4.29^{+0.26}_{-3.4}\,$\ms. Assuming a stellar mass of $M_*= 0.75\pm0.04\,$\MSun\ they placed an upper limit on the absolute mass of the planets to be $m_b=17.9^{+1.4}_{-14}\,$\MEarth\  and $m_c=17.18^{+1.1}_{-14}\,$\MEarth. Their calculations suggested a possible 5:3 mean motion resonance (MMR) between the two planets that would result in an oscillation of their period ratio ($P_c/P_b$). We analysed the 14 HARPS RVs and 7 publicly available CORALIE RVs alongside our ESPRESSO measurements, but due to their higher uncertainties, low number and sparseness, they do not improve our results. Hence, they were excluded from the final analysis. The HARPS and CORALIE measurements have a median uncertainty of 1.7 m/s and 6.0 m/s, respectively, while the peak-to-peak amplitude of the variation of the measurements is $19.18$ m/s and $33$ m/s, respectively.

Later, HD~23472 was observed with the Carnegie Planet Finder Spectrograph (PFS) mounted on the Magellan II telescope \citep{Teske2021}. They obtained 64 observations, allowing them to constrain the masses of the two outermost planets. The median of the uncertainty in the RV measurements is 0.63 m/s and the RV show a peak to peak amplitude of 12.10 m/s. Using their first method, which they refer as the 'juliet' method (without a correction for the activity), they found that the RV semi-amplitude for planet b is $K_b = 3.39 \pm 0.27\,$\ms, which corresponds to $m_b=11.7\pm 1.3\,$\MEarth,\ and for planet c it is $K_c = 1.18 \pm 0.30\,$\ms,  which corresponds to $m_c = 4.85 \pm 1.28\, $\MEarth. This is based on an assumed stellar mass of $M_*= 0.780\pm0.089\,$\MSun. They obtained slightly different results for the RV semi-amplitude, $K_b = 2.99 \pm 0.39\,$\ms and $K_c = 1.39 \pm 0.48\,$\ms using their second method, called 'radvel' method. This second method included the correction of stellar activity using a Gaussian Process (GP). We included the PFS measurements in our final analysis because they help to constrain the planetary RV signatures.

\subsection{TESS observations}

TESS observed HD~23472 (TIC 425997655, TOI-174) in five sectors (1,2,3,4,11) at a 2 minute cadence and four sectors at a 2 minute-and-20-second cadence (29,30,31,34). The non-continuous observations span is $\sim$ 900 days. We downloaded light curves computed by the TESS pipeline \citep{Jenkins2016} from the Mikulski Archive for Space Telescopes (MAST) \footnote{https://mast.stsci.edu/portal/Mashup/Clients/Mast/Portal.html}. We used the pre-search data-conditioned simple aperture photometry PDCSAP light curve \citep{Smith2012, Stumpe2012, Stumpe2014}, which corrects for the systematics of the light curves by removing trends that are common to all stars in the same CCD. We removed points with a quality flag other than zero, as well as points that deviated more than $5\, \sigma$ from a smooth version of the light curve. The individual sector light curves were normalised separately before being combined to produce the final light curve.

 Four planet candidates in HD~23472 were detected by the TESS Science Processing Operations Center (SPOC) using a wavelet-based adaptive matched filter \citep{Jenkins2002, Jenkins2010,Jenkins2020}. The data validation reports \citep{Twicken2018, Li2019} were reviewed by the TESS Science Office (TSO) and issued alerts in May 2019 \citep{Guerrero2021}. The four candidates have orbital periods of $P_b= 17.7$ days,  $P_c=29.8$ days, $P_d =3.98$ days, and $P_f =12.2$ days as well as radii $R_b  \sim 1.9\,$\REarth,  $R_c  \sim 2.1\,$ \REarth,  $R_d  \sim 0.8\,$ \REarth,\ and $R_f  \sim 1.2\,$ \REarth. A subsequent search of sectors 1-34 by the SPOC revealed a 5th planetary signature at a period of 7.908 days which  was alerted by the TSO on 21 October 2021. The TESS light curve shows no clear rotational modulation variability, and the GLS of the light curve shows no peak at the $40$ days activity signal seen in the RVs, neither half nor double of this period.

\section{Data analysis}
\label{analysis}
\subsection{Stellar parameters}

\label{stellarparameters}

\begin{table*}[h!]
\caption{Stellar parameters of HD~23472. \label{stellarp}}
\begin{center}
\begin{tabular}{l  c }
\hline
\hline
Parameter & Value and uncertainty\\
\hline
\textsc{ra}$^{\textsc{gaia-crf2}}$  [hh:mm:ss.ssss]  & 02:18:38.85 \\
\textsc{dec}$^{\textsc{gaia-crf2}}$ [dd:mm:ss.ss]    & -62: 46:02.18\\
B mag$^{\bullet}$                                               & $10.80  \pm 0.05$ \\ 
G mag  & 9.3899 $\pm$ 0.0028\\
V mag$^{\bullet}$                                                & $9.72 \pm 0.03$ \\ 
K mag$^{\bullet}$                                          & $7.21 \pm 0.023$ \\ 
H mag $^{\bullet}$                                        & $7.347 \pm 0.029$ \\ 
J mag$^{\bullet}$                           & $7.865 \pm 0.024$ \\
Effective temperature \teff\ [K]  & 4684  $\pm$ 99 \\
Surface gravity \logg\ [g cm$^{-2}$]  &  4.16 $\pm$ 0.24\\
Surface gravity \logg\  \footnotemark[1]  [g cm$^{-2}$]  &  4.53 $\pm$ 0.08\\
microturbulence [m/s] & 0.25$\pm$0.49\\
Iron abundance [Fe/H] [dex]  &  -0.20 $\pm$ 0.05\\
Magnesium abundance [Mg/H] [dex]  &  -0.19 $\pm$ 0.09\\
Silicon abundance [Si/H] [dex]  &  -0.18 $\pm$ 0.08\\
$\log R’_\text{HK}$   & $-4.9969 \pm 0.0002$ \\
vsin i & $1.45 \pm 0.17$ \\
Spectral type & K4V\\
Parallax*  $p$ [mas] & 25.581 $\pm$ 0.013\\
Distance to Earth* $d$ [pc] & 39.080  $\pm$ 0.019 \\
Stellar mass $M_{\star}$ [M$_\odot$] &0.67 $\pm$ 0.03 \\
Stellar radius $R_{\star}$ [R$_\odot$] & 0.71  $\pm$ 0.02\\
Stellar density $\rho_{\star}$  [$\rho_\odot$] & 1.88  $\pm$ 0.18\\
Stellar luminosity $L_{\star} $[L$_\odot$]  &0.237$\pm$  0.015 \\ 
\hline
\hline
\end{tabular}

*Parallax from \textit{Gaia} EDR3  \citep{Gaia2021} using the formulation of \citet{Lindegren2021}. Distance from \citet{Bailer-Jones2021}. \\
$^{\bullet}$  B and V magnitudes from \citep{Hog2000}  and K, H and J from \citep{2MASS} \\
\footnotemark[1] trigonometric surface gravity derived using the GAIA eDR3 parallaxes \citep{Gaia2021} 
\end{center}
\end{table*}

We combined the individual S1D ESPRESSO spectra after correcting for their radial velocities. The combined spectrum was then used to derived the stellar atmospheric parameters ($T_{\mathrm{eff}}$, $\log g$, microturbulence, [Fe/H]) and the respective uncertainties using ARES+MOOG, following the same methodology described in \citet[][]{Sousa2021, Sousa2014, Santos2013}. The analysis starts with the measurement of the equivalent widths (EW) of iron lines using the ARES code\footnote{The last version of ARES code (ARES v2) can be downloaded at http://www.astro.up.pt/$\sim$sousasag/ares} \citep{Sousa2007, Sousa2015}. We used a minimisation process to find ionisation and excitation equilibrium and converge to the best set of spectroscopic parameters. This process makes use of a grid of Kurucz model atmospheres \citep{Kurucz1993} and the radiative transfer code MOOG \citep{Sneden1973}. The line-list used for this analysis was taken from \citet[][]{Tsantaki2013}, which is more reliable for stars with effective temperature below 5200 K. The values derived for the temperature, log g, [Fe/H],  and microturbulence  are given in Table~\ref{stellarp}. Following the same methodology as described in \citet[][]{Sousa2021}, we used the distance derived from the GAIA eDR3 parallaxes \citep{Gaia2021} and estimated the trigonometric surface gravity to be 4.53 $\pm$ 0.06 dex and we adopted this value.

The stellar abundances of Mg and Si were derived using the classical curve-of-growth analysis method \citep{Griffin1967,Adibekyan2012} and assuming local thermodynamic equilibrium. We used the same tools and models as for the stellar parameter determination. Although the EWs of the spectral lines were automatically measured with ARES, we performed a careful visual inspection of these measurements to ensure that systematic noise sources, such as cosmic rays, do not affect the results. For the derivation of the abundance values, we followed the methods described in, for instance, \citet[][]{Adibekyan2012, Adibekyan2015}. The mean $\log R’_\text{HK}$ was derived by first extracting the $S_{Ca\,II}$ index from the co-added spectra with \verb+ACTIN+\footnote{\url{https://github.com/gomesdasilva/ACTIN}} \citep{GomesdaSilva2018}, converting it to the $S_\text{MW}$ scale using the calibration for ESPRESSO in \verb+pyrhk+\footnote{\url{https://github.com/gomesdasilva/pyrhk}} and calibrated to $R’_\text{HK}$ via the methodology described in \citet{GomesdaSilva2021}.

To determine the mass and radius of HD~23472, we used the Bayesian tool PARAM \citep{PARAM,PARAM2,PARAM3}, in which a set of observed quantities (namely, $T_{\rm eff}$, [Fe/H], luminosity) are matched to a well-sampled grid of stellar evolutionary tracks. 
The stellar luminosity is inferred from the 2MASS $K_s$ band \citep{2MASS}, corrected by distance (Gaia parallax) and the bolometric correction. The latter is estimated by YBC  \citep{YBC} that interpolate the observed effective temperature, \logg, and metallicity within a series of spectra libraries. Although extinction was considered in the calculation (estimated by STILISM \citealt{STILISM1,sTILISM2}), due to the proximity of the star, its impact on the luminosity is negligible with respect to overall statistical error (0.4\% against 15\%).  
As in \citet{PARAM2}, the grid of stellar evolutionary tracks and isochrones is from the PARSEC\footnote{\url{http://stev.oapd.inaf.it/cgi-bin/cmd}} code \citep[][]{Bressan2012}. 

For comparison, we computed the stellar parameters using the {\sc SteParSyn} code\footnote{\url{https://github.com/hmtabernero/SteParSyn/}} \citep{tab22}. The code implements the spectral synthesis method with an MCMC sampler to retrieve the stellar atmospheric parameters. We employed a grid of synthetic spectra computed
with the Turbospectrum \citep{ple12} code alongside MARCS
stellar atmospheric models \citep{gus08} and the atomic
and molecular data of the Gaia-ESO line list \citep{hei21}. We employed a set of \ion{Fe}{i,ii} lines that are well suited for the analysis of FGKM stars \citep{tab22}.  
In addition, STEPARSYN allowed us to compute the following stellar atmospheric parameters: $T_{\rm eff}$~$=$~4825~$\pm$~120~K, $\log{g}$~$=$~4.70~$\pm$~0.15 dex, [Fe/H]~$=$~$-0.13$~$\pm$~0.1~dex, including systematic errors.  
Using these stellar parameters, we used PARAM to derive a stellar mass of $0.72\pm 0.04\,$ M$_\odot$ and a stellar radius of $ 0.71\pm 0.03\,$ R$_\odot$.
These values are within $1~\sigma$ of the values derived with ARES+MOOG. We adopted the values of ARES+MOOG, which is the method used for most of the planetary systems analysed by the ESPRESSO GTO.

Assuming the relation between $\log R’_\text{HK}$ and the stellar rotation period derived by \citep{Mascareno2015} we estimated the stellar rotation to be $P_{rot}= 44 \pm 7$days. From the \vsini\ and the stellar radius, we can estimated the rotation period of the star to be  $\frac{25.1 \pm 3.0}{sin i}\,$days. Hence, to match the rotation period, we would need a stellar inclination of $40_{-6.8}^{+9.4}$ degrees.

 \subsection{Light curve analysis: A new planet}
 
 We performed our own transit search for additional transiting planets in the system. We used a method similar to that of  \citet{Barros2016}, which was further optimised for multi-planetary systems search. We began by applying a spline filter with breakpoints every 0.5 days to correct long-term variability in each sector.  Then we performed a transit search using a box least squares algorithm (BLS) \citep{Kovacs2002}. We analysed the phase-folded light curve at the period corresponding to the highest peak of the BLS periodogram to assess whether it is a good transit candidate. In any case, the duration and epoch provided by the algorithm were used to remove all points within the possible transit and close to it -- two durations before and after the mid-transit time. Due to the relatively high uncertainty in the period and epoch derived by the BLS algorithm, a cut wider than the transit duration is required. Depending on the complexity of the system or the noise in the light curve, this process was repeated several times.
 
For HD~23472, the first transit signal found by our procedure is the $17.7\,$day period planet (planet b), the second is the $29.8\,$ day period planet (planet c), and the third is the $12.2$ day period planet (planet f). However, the fourth detection is a new transiting planet candidate with a period of $7.9$ days (planet e) and the fifth is the $3.98$ day period planet (planet d). The orbital period of the new planet is nearly the double of the period of the previously known candidate at $3.98$ days, hence, we conducted several tests to confirm that there are indeed two planets. We discovered that when only the first four sectors of TESS are used, the fourth planet detected is planet d, and the new candidate is not detected. In our case, this is due to the fact that planet e transits just before planet d, and if planet d is discovered first, all transits of planet e are removed from the light curve when we remove transits of planet d, given our wide window cut. However, if planet e is discovered first, some of the transits of planet d remain. This could have been the reason planet e had not been previously detected. All transiting planet candidates pass the standard false positive tests \citep{Barros2016}. There is: no significant difference between odd and even transits; no detectable secondary eclipse; no ellipsoidal modulation at the planet period; and the transit duration is consistent with the stellar density derived by spectroscopy. We also used the open-source Python package \texttt(FULMAR) \citep{FULMAR} which uses the \texttt{Transit Least-Squares} (TLS) algorithm \citep{HippkeTLS2019} to retrieve and analyse the light curve and independently confirm the detection of planet e. At the time of this writing, a new candidate was presented the EXOFOP-TESS database \citep{TESSexofop} ($P_{05}=7.9$ days), supporting our new detection.

For further analysis of the transit photometry, we removed points in the light curve that are more than 1.5 transit durations away from each of the planet transits. Each transit was normalised by a linear trend computed from the out-of-transit flux close to each transit. Due to the uncertainty of the BLS-derived period and epoch, we used the ephemerides derived from the first iteration's multi-transit fit to recut the light curve and renormalise the data for our final analysis.

 \section{Radial velocity and light curve modelling}
 \label{modelfit}
 The light curve and the RVs were analysed simultaneously using the LISA code \citep{Demangeon2018,Demangeon2021}, which uses the \project{RadVel} python package \citep{Fulton2018} to model the RV observations and a modified version of the batman transit model\footnote{The modified version of batman is available at https://github.com/odemangeon/batman. The modification prevents an error for very eccentric orbits.} \citep{Kreidberg2015} to model the transits. The system is parameterised by the systemic velocity (${v_0}$), stellar density ($\rho_{\star}$), two limb darkening parameters for the TESS bandpass corresponding to the quadratic limb darkening law, and for each planet, the semi-amplitude of the RV signal (K), the planetary period (P), the mid-transit time ($T_0$), and the products of the planetary eccentricity by the cosine and sine of the stellar argument of periastron $e\cos{\omega}$, $e\sin{\omega}$,  the planet-to-star radius ratio ($r_{p}/R_{\star}$), and the cosine of the orbital inclination ($\cos i$). We included an offset between the two RV data sets ($\Delta\mathrm{RV}_{\mathrm{PFS/ESPRESSO}}$) and for each data set we included one additive jitter parameter  ($\sigma_{\rv, \mathrm{ESPRESSO}}$, $\sigma_{\rv, \mathrm{PFS}}$, $\sigma_{\tess}$. Moreover, we modelled the stellar activity seen in the RVs using a Gaussian process with a quasi periodic Kernel of the form:
\begin{equation}
    K_{\rv}(t_{i}, t_{j}) = {A_{\rv}}^2 \exp\left[-\frac{(t_{i} - t_{j})^2}{2 {\tau_{\mathrm{decay}}^2}} -\frac{\sin^2\left(\frac{\pi}{P_{\mathrm{rot}}} \abs{t_i - t_j}\right)}{2 \gamma^2} \right],
 \label{kernel}
\end{equation}
 where $A_{\rv}$ is the amplitude of the activity signal, $\tau_{\mathrm{decay}}$  is the decay timescale, $P_\mathrm{rot}$  is the period of the activity signal usually related to the stellar rotation period  \citep[e.g.,][]{Barros2020}, and $\gamma$ is the periodic coherence scale \citep[e.g.][]{Grunblatt2015}.  The GP was implemented with  the Python package \texttt{george} \citep{ambikasaran2015}.
 
 We used uniform priors for the parameters: systemic velocity, RV offset, jitter, planet-to-star radius ratio, semi-amplitude of the RV signal, the hyper-parameters of the GP, and the impact parameter. We included a prior in the impact parameter instead of the inclination to ensure the planets are transiting and to help convergence. We used Gaussian priors for the stellar density given in Table~\ref{stellarp}, the planetary period, and the mid-transit time derived from the BLS analysis and the limb darkening parameters derived with the LDTK code \citep{Parviainen2015,Husser2013} for the TESS bandpass ( $u_{1} = 0.472 \pm 0.056$ and $u_{2} =  0.186 \pm 0.055$). We used a Jeffrey prior for the eccentricity and constrained the eccentricity to be smaller than 0.15. Numerical simulations (Section~\ref{dinamics}) show that for the system to be stable, the eccentricity of all planets must be less than 0.1. If we do not include an informative prior, the eccentricity is poorly constrained to be less than 0.3, resulting in unstable solutions in the majority of the explored parameter space. Therefore, we imposed a strong prior in the eccentricity to ensure that the stable region of the parameter space was correctly sampled.
 
 LISA employs the Bayesian inference framework \citep[e.g.][]{Gregory2005} for parameter inference by maximising the posterior probability density function.  It explores the parameter space with the affine-invariant Markov-chain Monte-Carlo ensemble sampler implemented in \emcee\ \citep{Goodman2010,Foreman-Mackey2013}. The number of walkers is set to 2.5 times the number of fitted parameters by default. To speed up convergence, the starting parameters are randomly drawn from the prior and a pre-minimisation is performed using the Nelder-Mead simplex algorithm \citep{Nelder1965} implemented in the Python package \texttt{scipy.optimize}. The chains were checked for convergence using Geweke test \citep{Geweke1992} and the burning-in part of the chain was removed before merging the chains.  The final clean chain was composed of 5\,310\,000 values. The best value for each parameter was derived from the median of the posterior distribution and the uncertainties from its 68\% confidence interval. More details on the LISA fitting procedure is available  in \citet{Demangeon2018, Demangeon2021}.

\begin{figure*} 
\centering 
\includegraphics[width=1.8\columnwidth]{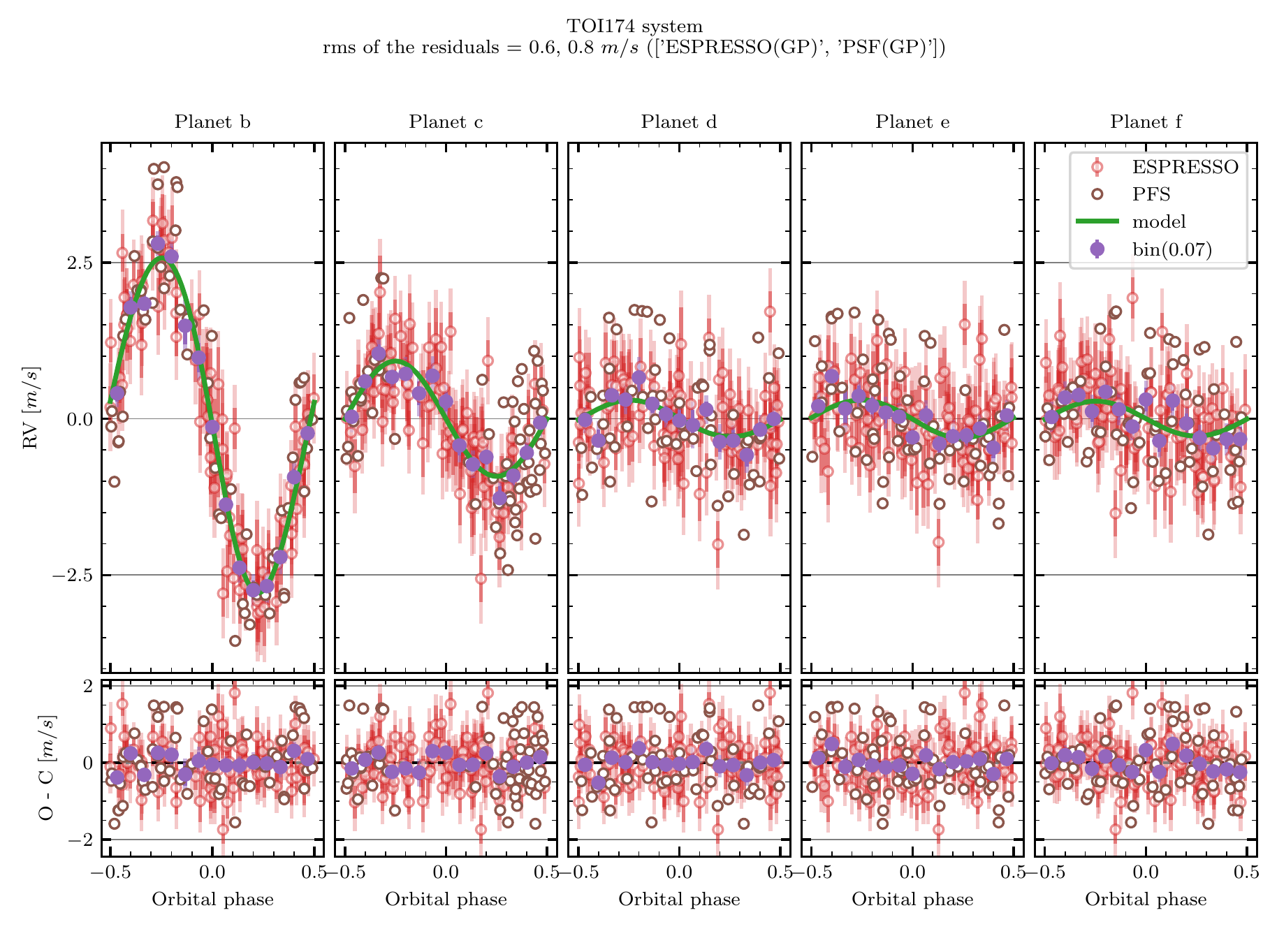}
\caption{Phase-folded RVs of the ESPRESSO (in red) and PFS (in brown) data in the periods of the HD~23472 system's five planets are shown in the top panel. The best model is shown in green. The RVs were corrected for systemic velocity, offset between the two instruments, and stellar activity with the fitted GP model. For clarity, we also show the binned RVs and, in addition, the uncertainties of the PFS data are not displayed, but they are slightly larger than the ESPRESSO ones, as mentioned in Section~\ref{observations}. The residuals relative to the best-fit model are shown in the bottom panel. \label{faseRVs}} 
\end{figure*}

\begin{figure*} 
\centering 
\includegraphics[width=2.0\columnwidth]{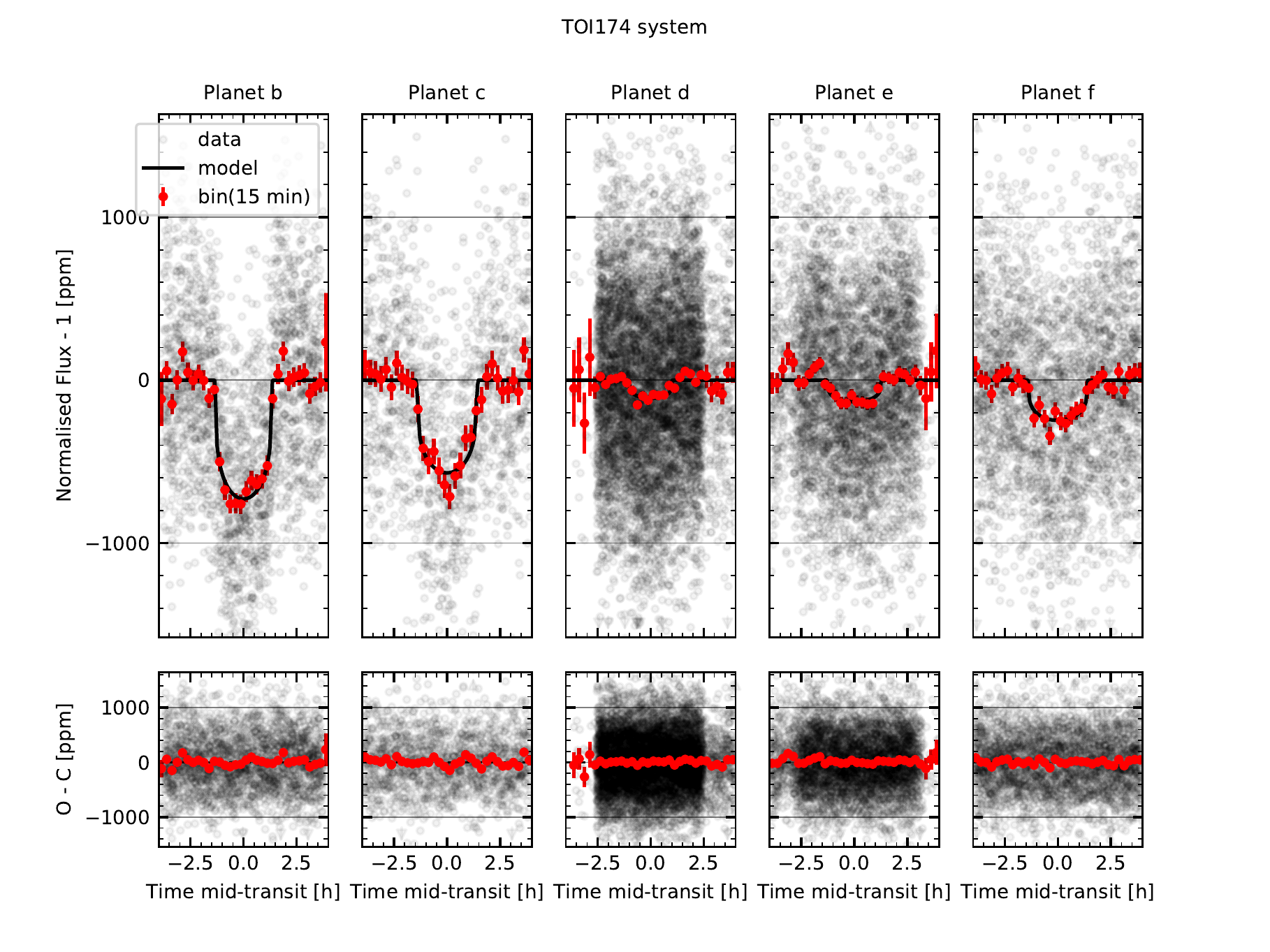}
\caption{Phase-folded transit light curves (grey dots) obtained by the TESS satellite (top panel). We overplot the 15 minute binned light curves and the corresponding uncertainties in red. We also oveplot the best fit model in black. The uncertainties of the unbined data were not displayed for clarity. Residuals relative to the best-fit model (bottom panel). \label{fasetransits}} 
\end{figure*}

Figure~\ref{faseRVs} shows the best model for the five planets overploted on the RV observations corrected for stellar activity with the fitted GP model,  while Figure~\ref{fasetransits} shows the best transit model overploted on the transit light curves.  The five planets' transits are clearly evident, but the RV signature is only clear for the two largest and longest period planets, b and c, which have already been confirmed.  

 \begin{figure*} 
\centering 
\includegraphics[width=2.0\columnwidth]{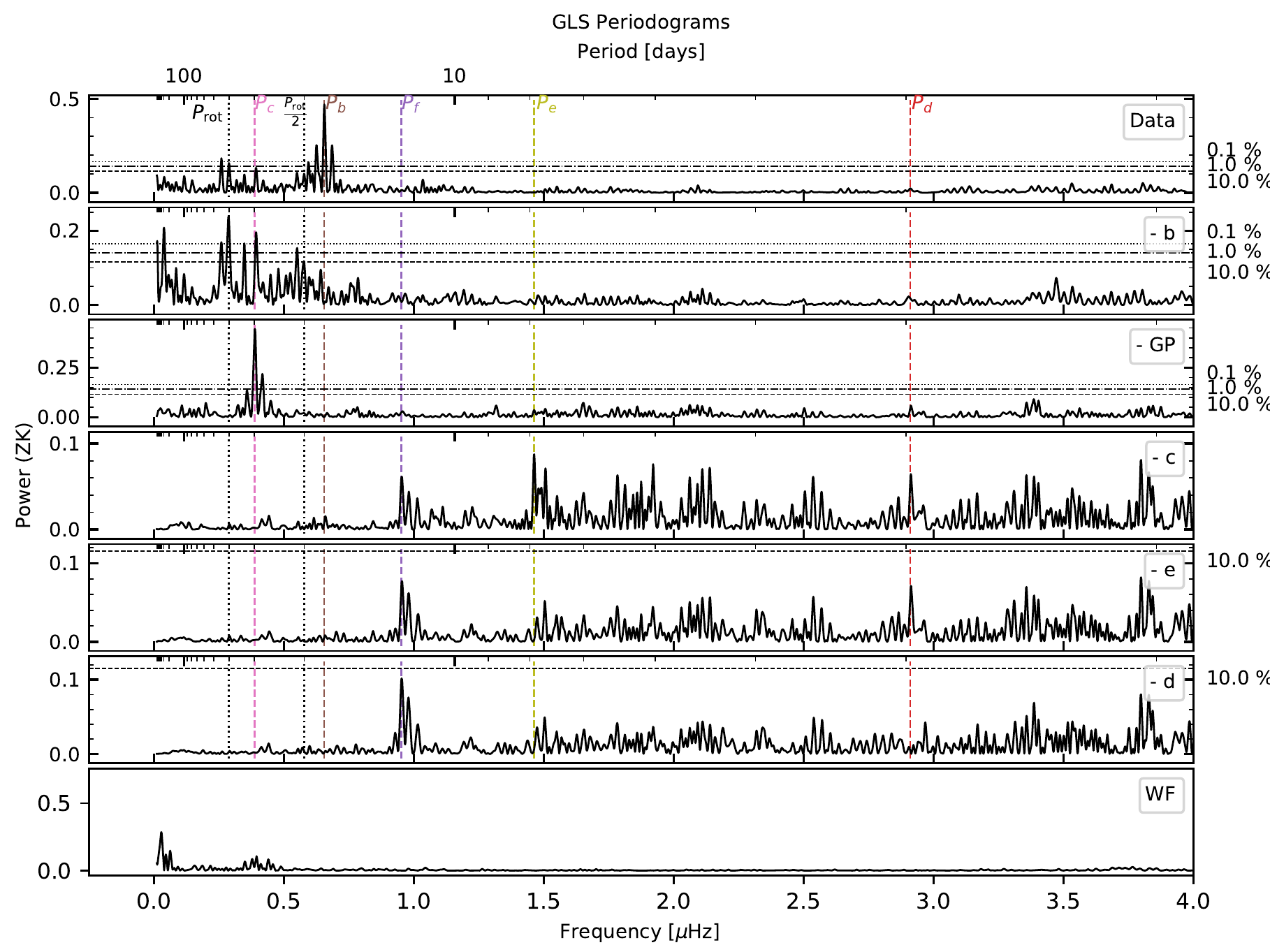}
\caption{Iterative GLS for the five planet model. The top panel shows the GLS of the RVs (ESPRESSO and PFS after correcting for the instrument offset). The subsequent panels show the previous row minus the model of planet b, model of the GP, model of planet c, model of planet e, and model of planet d, respectively. The last panel shows the window function. The horizontal lines indicate the 10\%(dashed line), 1\%(dot-dashed line), and 0.1\% (dotted line) FAP levels calculated following \citet{Zechmeister2009}. \label{GLSiteractive}} 
\end{figure*}

 Table~\ref{syspar} displays the  best-fit model parameters. The system has three inner planets that are similar in size to the Earth but have a lower mass. The two innermost ones are smaller and more massive, so they have higher densities than the three outermost ones. The density of the middle planet f is lower than that of the Earth. 
 The two outermost planets have the largest radii, $\sim 2\, $\REarth, with planet b having more than twice the mass of planet c and hence being denser. As the figures show, we obtained a good precision, that is, better than $8.7 \%$, for the radius measurement of all five planets in the system. Our precision, however, is lower for the masses. The two outermost planets are detected at more than $4\sigma$, planets d and e have a hint of detection at $2.7\,\sigma$, and planet f's semi-amplitude of the RV is measured at $1.9\sigma$. As a result, for the three smaller inner planets, particularly planet f, we advise caution in interpreting the planet's composition.  The relative precisions in the derived semi-amplitude is 9\%, 25\%, 37\%, 38\%, and 55\% for planets b, c, d, e, and f, respectively.

    Comparing our results with previous estimates of the RV semi-amplitude for planet b and planet c derived using the PFS data \citep{Teske2021}, we find that our results are in better agreement with the second (RadVel) method, which includes a GP to correct stellar activity than with the first (juliet) method, which, for this target, did not include the correction of stellar activity. As can be seen in Figure~\ref{RV_modelzoom}, the start of the ESPRESSO observations coincides with the end of the PFS observations, and there is no substantial difference in the amplitude of the stellar variability between the earlier season of PFS observations and the later season of ESPRESSO observations. Furthermore, a GLS of the PFS observations reveals a clear peak at the stellar rotation period of the star $~\sim 40.5\,$ days. Therefore, the new ESPRESSO observations allow to improve the precision and the accuracy of the mass measurements of the outermost super-Earths in the  HD~23472, as well as to measure the mass and density of planets d and e and derive a strong upper limit on the mass of planet f.

  \subsection{Exploring the correction of the stellar activity}

  The ESPRESSO DRS pipeline, as mentioned in Section~\ref{espressoobservations}, provides time series for several activity indicators (FWHM, BIS, S-index, NaD, $H_\alpha$, contrast). These activity indicators assess the distortions and depth of stellar lines, which serve as proxies for stellar activity. They may also indicate the presence of blending caused by another star or an eclipsing binary. Because they are not affected by the presence of planets, they are ideal for distinguishing the periodicities caused by stellar activity. Figure~\ref{indicators} compares the time series and GLS periodograms of the RVs and activity indicators. The activity indicators show several statistically significant peaks at long periods, but none at the planet's periods. The third strongest peak in the RVs corresponds to significant peaks in all activity indicators, a clear sign of stellar activity. We identify this as the rotation period of the star, which is approximately 40 days. This is in agreement with the values of the rotation period of the star estimated in Section~\ref{stellarparameters} from $\log R’_\text{HK}$. As mentioned above, the TESS light curve shows no clear signs of rotational modulation and no periodicity at 40 days. Hence, the light curve was not used to constrain the stellar activity signal.

Using the information contained in the activity indicators can assist in correcting for stellar activity. This was accomplished by modelling the activity indicators with a GP using the same Kernel as the RVs (equation \ref{kernel}). The RVs and one activity indicator are fitted simultaneously. The GPs modelling the RV and the activity indicator are independent but the value of their hyper-parameters are assumed to be the same except for the amplitude. For the activity indicators, we assumed that the mean function is a constant value that we also fit. We repeated this procedure using the activity indicators FWHM, $H_\alpha$, and contrast, individually, and we also performed the fit using the two activity indicators, FWHM and $H_\alpha$, simultaneously. We found that there was no significant improvement in the activity modelling when using the activity indicators. The fitted hyper-parameters with and without activity indicators have similar values and uncertainties. We also found no improvements in the RVs' residuals. This is most likely due to two factors. First, the RVs are well sampled and their precision allows for a good constraint on the activity model's hyper-parameters. This is supported by the well constrained hyper-parameters' posteriors shown in Figure~\ref{cornerhyper}. Second, the power spectra of the activity indicators and the RVs differ slightly, which could result in different best fit hyper-parameter values. For example, there may be periodicities in the activity indicators that are not present in the RVs, which could affect the hyper-parameter values. This could be related to the stellar activity of HD~23472 being plage-dominated rather than spot-dominated. Stars with spot-dominated stellar activity have been found to have RVs that are better correlated with the activity indicators. 

A better understanding of the correlation and usefulness of RV activity indicators is of extreme importance to obtain accurate masses of exoplanets. Therefore, we decided to test the best-fit values of individual activity indicators. We performed a simultaneous fit of RV, FWHM, $H_\alpha$, and contrast. In this case, we did not simultaneously fit the transit light curve, but instead we included Gaussian priors for the period and the transit epoch of the five planets. For the activity indicators, we used a constant mean function whose value we fit as before as well a GP to model the covariance matrix using the same RV model as presented above. In this case the only hyper-parameter shared between the activity models for the four data sets is the rotation period. Although the amplitude of the activity models for the activity indicators is expected to be different, the values of the decay timescale ($\tau_{\mathrm{decay}}$) and the periodic coherence scale ($\gamma$) are typically considered to be the same \citep{Demangeon2021, Mascareno2020}. The values we derived for the decay timescale and the coherence scale are given in Table~\ref{fitind}. The rotation period value was found to be $P_{rot}= 39.9^{+1}_{-0.91}$ days, which is the same as the value for the period derived if we fit only the RVs (see Table~\ref{syspar} ).

We found that the derived values for the decay timescale and the coherence scale from the FWHM, $H_\alpha$, and contrast are significantly different from the values derived from the RVs, in particular for the coherence scale. The coherence timescale derived for the activity indicators is more similar to each other than any of them is to the RVs. The derived value of the decay timescale for the FWHM is the closest to the derived value for the RVs agreeing within $1 \sigma$. This difference in the hyper-parameters of the activity models for the activity indicators is probably the cause for the lack of improvement in including the activity indicators in the global RV fit.  

With our new insight into how the hyper-parameters vary for different activity indicators we tested the correction of stellar activity including the FWHM in our systems fit of the RV and the photometry. However, in this new test, we assume only $P_{rot}$ and $\tau_{\mathrm{decay}}$ to be the same for both the RV and the FWHM activity model and $A_{\rv}$ and $\gamma$ are independent for the RV and the FWHM activity model. The derived values for rotation period is similar to our previous results with a similar error $P_{rot}= 39.5^{+0.92}_{-0.82}$ days and the derived the decay timescale smaller and with half of the uncertainty of the previous result. We also confirmed that the values of coherence scale for the RV and FWHM are significantly different. These improvements, however, did not affect the derived planetary parameters. All the planetary parameters are within $0.1 \sigma$ of those previously derived and have similar errors. There was no reduction of the jitter derived for the RVs or the rms of the residuals. Given that there was no improvement, we conclude that the added complexity of the model is not justified. Further insights could be obtained by probing whether this behaviour is shared by other stars. A better understanding of the correlation and utility of RV activity indicators is essential to obtain accurate exoplanet masses and more research in this area is required.

\begin{table}[h!]
\caption{GP hyper-parameter estimates for the RV and the activity indicators \label{fitind}}
\begin{center}
\begin{tabular}{l  c c }
\hline
\hline
Indicator &  $\tau_{\mathrm{decay}}$  &$\gamma$\\
 &  days  & \\
\hline
\\[-8pt]
RV              & $1789_{-610}^{+860}$ &$0.368_{-0.057}^{+0.074}$ \\
FWHM        & $1110_{-331}^{+395}$ & $0.87_{-0.14}^{+0.19}$ \\
$H_\alpha$ & $82_{-37}^{+860}$ & $2.0_{-1.2}^{+1.9}$ \\
contrast       &$681_{-543}^{+670}$  &$1.4_{-0.51}^{+1.2}$ \\ 
\\[-8pt]
\hline
\hline
\end{tabular}
\end{center}
\end{table}

With a $\log R’_\text{HK} \sim -5$ dex, TOI-174 is a chromospherically inactive K dwarf, typical of old main sequence (MS) stars \citep{Henry1996}.
Inactive K dwarfs have higher activity variability than inactive FG dwarf stars \citep{GomesdaSilva2021} but have lower induced RV variability \citep{Lovis2011}.
The measured value of the amplitude of the RV variation for HD~23472 ( ${A_{\rv}} \sim 2$\ms ) is in agreement with the typical RV variability ($\leq 2.5 m/s$) of MS stars with similar mass and activity levels \citep{Luhn2020}. The measured rotation of $\sim$40 days, obtained independently from the activity-rotation relation and activity time series, is at the upper envelope of the rotation periods distribution of stars with similar effective temperatures and is typical of an old K dwarfs with ages of $\geq$4.5 Gyrs \citep{McQuillan2014, Angus2020}.

 \subsection{ESPRESSO performance}
 \label{performance}
 Because ESPRESSO is a new instrument, it is interesting to compare the results of our final adopted model combining the ESPRESSO and PFS data with the results of the analysis using only ESPRESSO data. Table~\ref{sysparesp} displays the most relevant parameters for the ESPRESSO + TESS fit with the same priors as in the final model (Table~\ref{syspar}).

All of the results are found to be within $1 \sigma $ of the combined ESPRESSO and PFS results. However, the amplitudes of the RV variation measured using only the ESPRESSO have a slightly lower level of precision, which is expected due to the decrease in the number of RV measurements. When the PFS data are included, the precision increases by $12-9\%, 35-25\%, 56-37\%, 57-38\%,$ and $ 56-55\% $ for planets b, c, d, e, and f, respectively. The mean of the residuals after removing the GP is 0.796 m/s for the PFS data and 0.555 m/s for the ESPRESSO data and the jitter parameter for the PFS data is also slightly larger than for the ESPRESSO data as can be seen in Table~\ref{syspar}.

\section{Dynamics and orbital architecture}
 \subsection{Dynamical stability and refinement of eccentricities}
\label{dinamics}

In orbital period ascending order, the period ratios of the HD~23472 planets are as follows: $P_e/P_d =1.99$, $P_f/P_e =1.54$, $P_b/P_f =1.45$, and $P_c/P_b =1.68$. These pairs are somewhat close to mean motion resonances (MMRs) where $ P_{out}/P_{in} \approx (k+q)/k$. The three innermost pairs are closer to first-order MMRs (q=1, k=1 or 2), while the outermost pair is closer to a second-order MMRs (q=2, k=3). In case of multi-planetary systems close to several two-planet MMRs, similar super-periods amongst successive pairs $P=1/((k+q)/P_{out}-k/P_{in})$ \citep{Lithwick2012} have been shown to indicate the presence of a Laplace relation between successive planet triplets like in Kepler-60, Kepler-80, Kepler-223, K2-138, Trappist-1, and TOI-178 \citep[][]{Jontof-Hutter2016,MacDonald2016,Mills2016,Lopez2019,Luger2017,Leleu2021}. This relation is not satisfied by any of the successive pairs in HD~23472. The proximity to the two-planet MMRs system, on the other hand, suggests that it may have formed in such a chain in the protoplanetary disc, but was later disrupted \citep{Terquem2007,Goldreich2014,PuWu2015,Batygin2015, Deck2015b,Izidoro2017}. In particular, the inner pair displays a period ratio less than the exact commensurability ($P_e/P_d<2$). This is not what we would expect for a pair of planets that are initially inside a first order MMR and evolve smoothly through tides (typically effective for orbital periods less than 10 days): for such a pair, we would expect tidal evolution to gradually increase the period ratio toward $P_e/P_d>2$. The period ratio is less than the exact commensurability, which could be due to a disruption early in the system's history, possibly near the end of the protoplanetary disc phase or before tides had a chance to come into play.

 \begin{figure*}[!ht]
\begin{center}
\includegraphics[width=0.3\textwidth]{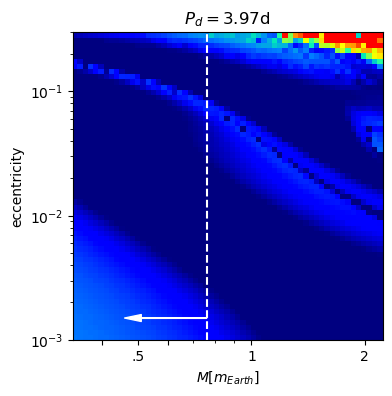}
\includegraphics[width=0.3\textwidth]{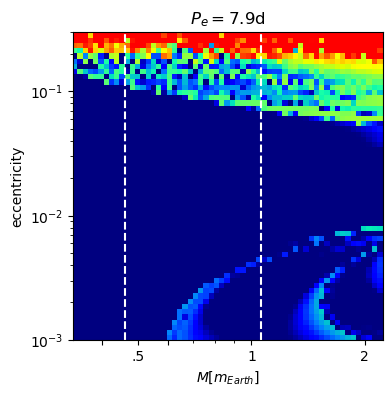}
\includegraphics[width=0.3\textwidth]{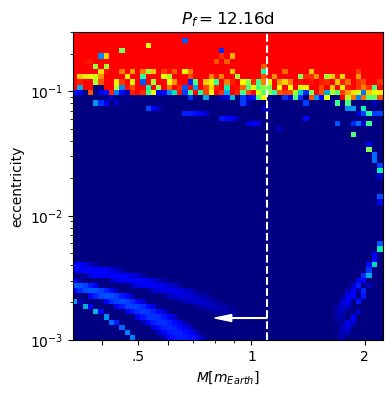}\\
\includegraphics[width=0.3\textwidth]{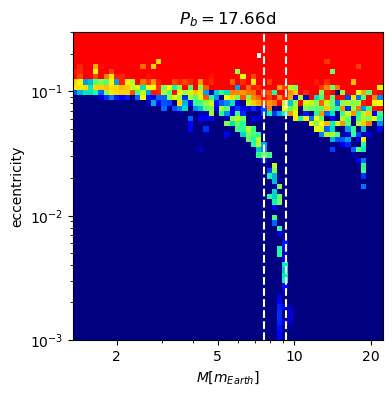}
\includegraphics[width=0.3\textwidth]{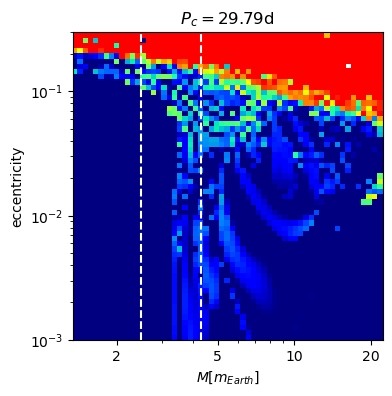}
\includegraphics[width=0.07\textwidth]{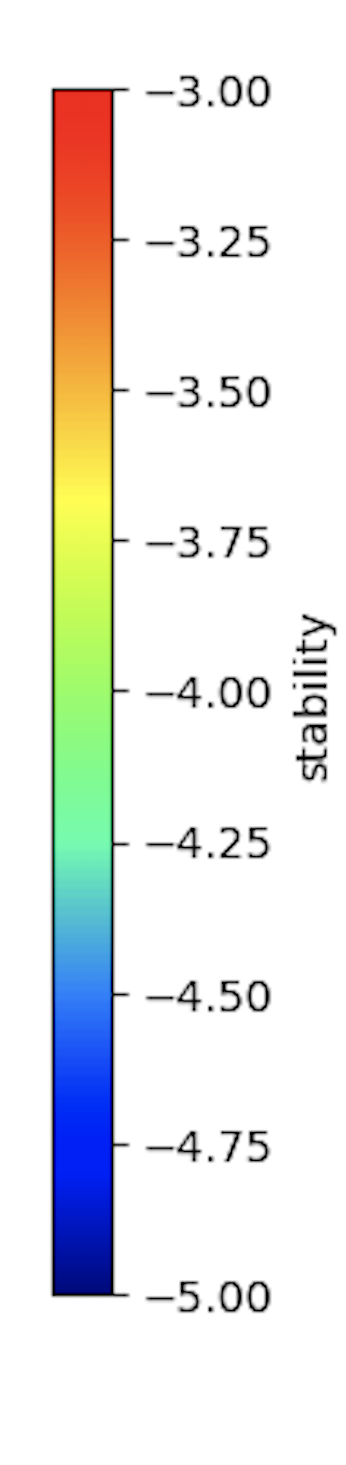}
\caption{\label{fig:stabMe} Stability maps as function of the mass and eccentricity of each of the planet of the HD~23472 system. The stability criterion is given in Equation~\ref{eq:diffusion}. Blue is short-term stable while red is short-term unstable. For each map, the parameters of the rest of the system are set to the nominal value given in Table~\ref{stellarp}, except for the eccentricities that are set to zero. The white dashed lines show the .16-.84 quantiles of the posterior of the mass of each planet. Only the .84 quantile is shown for planets d and f. }
\end{center}
\end{figure*}

\begin{table}[h!]
\caption{Dynamics-corrected eccentricity. \label{eccstab} }
\begin{center}
\begin{tabular}{ c c }
\hline
\hline
planet &  eccentricity \\
\hline
\\[-8pt]
b & $0.042_{-0.026}^{+0.034}$\\
c  & $0.056_{-0.039}^{+0.051}$\\
d & $0.069_{-0.046}^{+0.053}$\\
e & $0.056_{-0.038}^{+0.054}$\\
f  &$0.048_{-0.034}^{+0.044}$\\
\\[-8pt]
\hline
\hline
\end{tabular}
\end{center}
\end{table}

We first performed a short-term stability analysis of the posterior samples in the case where the eccentricity of the planets is left free. 
For this analysis, we used the frequency analysis stability index, which is based on the diffusion of the system's main frequencies and is defined as \citep{Laskar1990,Laskar1993}:
\begin{equation}
\log_{10} \left\vert \frac{n^{(1)} - n^{(2)}}{n^{(1)}}\right\vert  \, ,
\label{eq:diffusion}
\end{equation}
where $n^{(1)}$ and $n^{(2)}$ are the proper mean motion of the planets, computed over the first half and the second half of the integration. We found that a majority the posterior samples in the case where the eccentricity of the planets is left free is unstable in the short term. Given the system's compactness, this might be due to the relatively large eccentricity values for the free-eccentricity solution. Because the eccentricities and some of the masses are highly uncertain, we explore these parameters to gain a better understanding of their impact on the system's stability. Figure \ref{fig:stabMe} displays stability maps varying the mass and eccentricity of each planet. The orbital parameters and masses of the other planets are set to their median value in each case, with the exception of eccentricities, which are set to zero. The colour code corresponds to the maximum of the stability index given in Equation \ref{eq:diffusion} for all planets, implying that the stability increases from red to blue \citep[for more details, see Section 4.1 of][]{Petit2018}. The maps show unstable structures that are hard to interpret in the mass-eccentricity plane, but are typically associated with the presence of nearby MMRs \citep[e.g.][]{Leleu2021}, the precise position and boundaries of which also depend on the masses and eccentricities \citep[see][]{HenLe1983}. A significant fraction of the 1$\sigma$ mass interval is stable for all planets, pointing to eccentricities as the main parameters indicative of a system's stability. Another common trend is the appearance of instabilities for eccentricities typically greater than $\sim 0.1$.

We performed a second MCMC, exploring the eccentricities in the [0,0.15] range for all planets, because setting upper limits on each eccentricity while forcing the other eccentricities to be equal to zero is not satisfactory. We then cut the posterior in two parts using the method presented in \cite{Stalport2022}: we set the stability index threshold to -4 (orange in Fig. \ref{fig:stabMe}), removing, from the posterior, all trajectories that are short-term unstable  ($37\%$  of the trajectories were discarded as such). When examining the new, short-term stable posterior, the $0.16$ and $0.84$ quantiles of most parameters remain roughly unchanged, with the exception of some eccentricities. The estimated eccentricities, when accounting for the stability of the system are given in Table~\ref{eccstab}.
 The posterior of the eccentricities of HD~23472~b and HD~23472~f shifted towards a lower value, which is required to ensure the system's stability.

What is noteworthy, according to our final result, the three inner pairs are outside of MMR. The outer pair, on the other hand, is quite close to the 5:3 MMR, and additional monitoring could confirm its resonant state. The transit timing variations (TTVs) between resonant pairs are strongly dependent on eccentricity, which is not well constrained in our case. Assuming zero eccentricity, we estimate the peak-to-peak amplitude of TTVs for planet b to be 4 minutes and for planet c to be 6 minutes. We used the same method as \citet{Barros2022} to calculate the observed TTVs from the TESS light curve and found no significant TTVs. Our estimated errors, however, have the same amplitude as the expected TTVs, 6 minutes for planet b and 7 minutes for planet c. Therefore, we cannot rule out the existence of TTVs. Photodynamical modelling the system \citep[e.g.][]{Barros2015} could improve the precision of the measured TTVs and help constrain the system parameters, but this approach is beyond the scope of this paper.

 \subsection{Architecture}

\begin{table*}[!htb]
\caption{Peas in a pod statistics in the system HD~23472 \label{peas} }
\begin{tabular}{p{0.9\columnwidth}p{0.5\columnwidth}} 
\hline \noalign{\smallskip} 
Metric from the  HD~23472 system & W18 distribution \\      
\hline \noalign{\smallskip}
$R_e / R_d = 1.10 \pm 0.12$ & $1.14 \pm 0.63$\\ \noalign{\smallskip}  
$R_f / R_e =  1.39  \pm  0.14$ &  \\ 
$R_b / R_f =  1.76 \pm 0.11$ &  \\ 
$R_c / R_b = 0.937 \pm 0.049 $ &  \\ 
\noalign{\smallskip} \hline \noalign{\smallskip}
$(P_f / P_e) / (P_e / P_d) = 0.773472 \pm  0.000022   $ & \multirow{2}{0.40\columnwidth}{$1.00 \pm 0.27$}\\  
$(P_b / P_f) / (P_f / P_e) =   0.944459 \pm  0.000020  $ & \\
$(P_c / P_b) / (P_b / P_f) = 1.161078 \pm   0.000016 $ & \\
\noalign{\smallskip} \hline \noalign{\smallskip}
$\Delta(b, c) = 13.30 \pm 0.45 $ & \multirow{4}{0.40\columnwidth}{Mode between 10 and 20 }\\
$\Delta(f, b) = 10.39 \pm  0.34$  & \\
$\Delta(e, f) = 22.3 \pm 3.0 $ & \\
$\Delta(d, e) = 36.9 \pm 3.9$ & \\
\noalign{\smallskip} \hline \noalign{\smallskip}
$T_{\textrm{eq}, d} - T_{\textrm{eq}, e} =185 \pm 32$ K & \multirow{2}{0.40\columnwidth}{$T_{\textrm{eq}, i} - T_{\textrm{eq}, i+1}$ positively correlated with $R_{i + 1} / R_{i}$} \\
$T_{\textrm{eq}, e} - T_{\textrm{eq}, f} = 96 \pm 28$ K & \\
$T_{\textrm{eq}, f} - T_{\textrm{eq}, b} = 79 \pm 21$ K & \\
$T_{\textrm{eq}, b} - T_{\textrm{eq}, c} = 80 \pm 18 $ K & \\
\noalign{\smallskip} \hline \hline           
\end{tabular}
\tablefoot{- $\Delta(i, j)$ is the separation in mutual Hill radii (see Eq. 5 in W18)\\
When the notation $x \pm y$ is used in the column " W18 distribution",  $x$ is the median of the observed distribution, and $y$ is its standard deviation.}
\end{table*}

A five planet system is an excellent test of the 'peas in the pod' trends reported by \citet{Weiss2018}.  We investigated whether HD~23472 follows the main trends presented in \citet{Weiss2018}, listed below in italic. Table~\ref{peas} gives the values of the "peas in the pod" metrics. Our main conclusions are  as follows:

 \textit{The size of adjacent planets is similar or the outer planet is larger than the inner planet}: The planets in HD~23472 have similar radii and the planetary radius increases outwards from the star for the first four planets, but decreases slightly for the planet with the longest period.
 \textit{The orbital spacing between planets is similar}:  This also occurs in HD~23472, although the orbital spacing increases slightly as one moves away from the star.
 \textit{Smaller planets are more packed than larger planets}: This is in contrast to HD~23472, where the larger outer planets are more densely packed than the inner ones (i.e. they have a smaller separation in units of mutual hill radius).
 \textit{The temperature difference correlates with the planet size ratios}: This occurs in the HD~23472 system as well.

Our high errors in terms of  mass prevent us from testing whether the planets are more similar in mass than in radius,  as found by  \citet{Otegi2022}. Additional RV observations to improve mass precision would be extremely helpful in testing this hypothesis.  HD~23472 follows the general trends of "peas in the pod" but it appears to be breaking the trends for the longest period planet. As a result, probing longer period planets in this system would be very beneficial.

\section{Internal structure}

\label{composition}

\subsection{Planetary composition}
\begin{figure}[!ht]
\begin{center}
\includegraphics[width=0.49\textwidth]{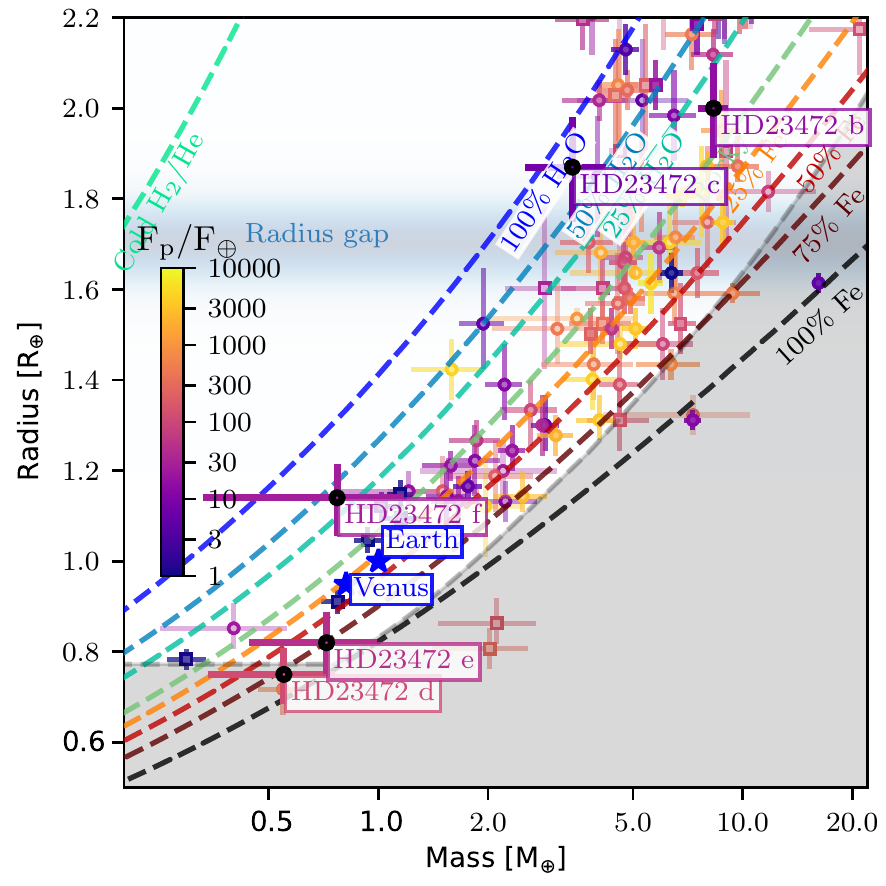}
\caption[]{ \label{figmassradius} HD~23472 planets in the context of other known transiting planets with measured mass and radius precision better than $50\,\%$. The exoplanets data have been extracted from the NASA exoplanet archive. Planets whose masses were determined using the RV technique are represented as circles, whereas planets whose masses were determined using transit timing variations are represented as squares. The intensity of the incident flux is indicated by the colour of the points.The planetary bulk density's relative precision is proportional to the transparency of the error bars. The planets of the solar system are illustrated as blue stars. We also show the mass-radius models of \citet{Zeng2016} as dashed lines. The radius gap \citep{Fulton2017} is shown as a shaded horizontal blue line and the maximum collision stripping of the mantle region is shown in grey. This graph was created using the mass radius diagram code\footnotemark .}
\end{center}
\end{figure}

Our analysis enabled us to achieve relative precisions in the densities of the five planets of $18\%, 31\%, 46\%, 46\%$, and $ 59\% $  for planets b, c, d, e, and f, respectively. Although this result is remarkable for such low-mass planets,  a higher level of precision would be necessary to uniquely characterise the composition of the planets especially the three smaller ones and we advise taking care in the interpretation of the results. Figure~\ref{figmassradius} displays the position of the five planets in the mass-radius diagram as well as the compositional models of \citet{Zeng2016} and the  radius gap \citep{Fulton2017}. The three inner planets are below the radius gap while the two outermost planets are above it. This could be caused by either irradiation shaping the system or core-powered evaporation. The two inner planets, d and e, are very dense, indicating that they likely contain more iron than Earth according to the models of  \citet{Zeng2016}.  Planets b, c, and f appear to have a significant amount of water or gas in their composition, with planet c having the most.  Moreover, the two inner planets are in a sparsely populated region of the parameter space near L\,98-59\,b, whose mass was recently measured with ESPRESSO  \citep{Demangeon2021} and Trappist-1\,h \citep{Gillon2017}, whose mass was measured with transit timing variations. Finding and characterising planets in this region of the parameter space would greatly improve our understanding of the composition of small exoplanets.

\footnotetext{The code is available online at \url{https://github.com/odemangeon/mass-radius_diagram}}.

To constrain the internal structure of the planets, we performed a Bayesian analysis following the method of \citet{Dorn2015, Dorn2017}. The same method was used to analyse other systems, including, L\,98-59\,b  \citep{Demangeon2021}, TOI-178 \citep{Leleu2021}m and Nu2 Lupi \citep{Delrez2021}. The model assumes the planet has four layers: an  iron and sulphur inner core, a silicate mantle ( made of Si, Mg and Fe), a water layer, and a gas layer (made of H and He). We used an improved equation of state for the water layer \citep{Haldemann2020} and an improved equation of state of the iron core \citep{Hakim2018} for our analysis.

The model includes two parts: a forward model that computes the planetary radius as a function of the internal
structure parameters and a Bayesian analysis that computes the posterior distribution of the internal structure parameters needed to fit the observed radii, masses, equilibrium temperatures, and stellar parameters. Since absolute planetary masses and radii are measured relative to the same host star, they are correlated. As a result, rather than fitting the absolute masses and radii, we fit the radius ratio and radial velocity semi-amplitudes of all planets in the system simultaneously.  The input parameters of our model are the stellar mass, stellar radius, stellar effective temperature, stellar age, stellar chemical abundances of Fe, Mg, and Si, and the planetary radial velocity semi-amplitudes, planetary radius ratio, and orbital periods. The fitted parameters are the mass fractions of core, mantle, water layer, and gas layer. Except for the mass of the gas layer, which is assumed to follow a uniform-in-log prior, we used uniform priors for these parameters. We note that the results we obtain are influenced by these priors to some extent. Assuming a uniform prior for the gas mass, for example, would result in a large mass fraction of gas and a smaller mass fraction of water. Finally, the mass fractions of the different layers add up to one, and water mass fractions greater than 50\% are excluded \citep{Thiabaud2014,Marboeuf2014}.

We ran a first set of models, assuming, as in other similar studies that the planets share the same Si/Mg/Fe molar ratio equal to the stellar one (see, however, \citet{Adibekyan2021}). In this case, it turns out to be impossible to fit the mass and radius of the two innermost planets (d and e), whose densities are too high to be reproduced by a core-mantle structure with stellar abundances. We then relaxed the assumption of stellar composition for these two planets, letting the Si/Fe and Mg/Fe ratio free, but assuming the Si/Mg ratio to be stellar. We note that this last assumption comes from the idea that in the case of the formation of a Mercury-like planet by a giant collision that would have removed part of the mantle, it is not expected that Si and Mg would fractionate \footnote{Since our model assumes that the mantle is homogeneous (i.e. the Si/Mg ratio is uniform) and there is no Si and Mg in the core, then if there is an impact that removed the mantle layers, the Si/Mg ratio is preserved.}. All other assumption regarding the forward model are kept the same. In this case, it is possible to reproduce the five planets in an accurate way. 

\begin{figure*}[!htb]
 \centering
 \includegraphics[width=0.45 \hsize]{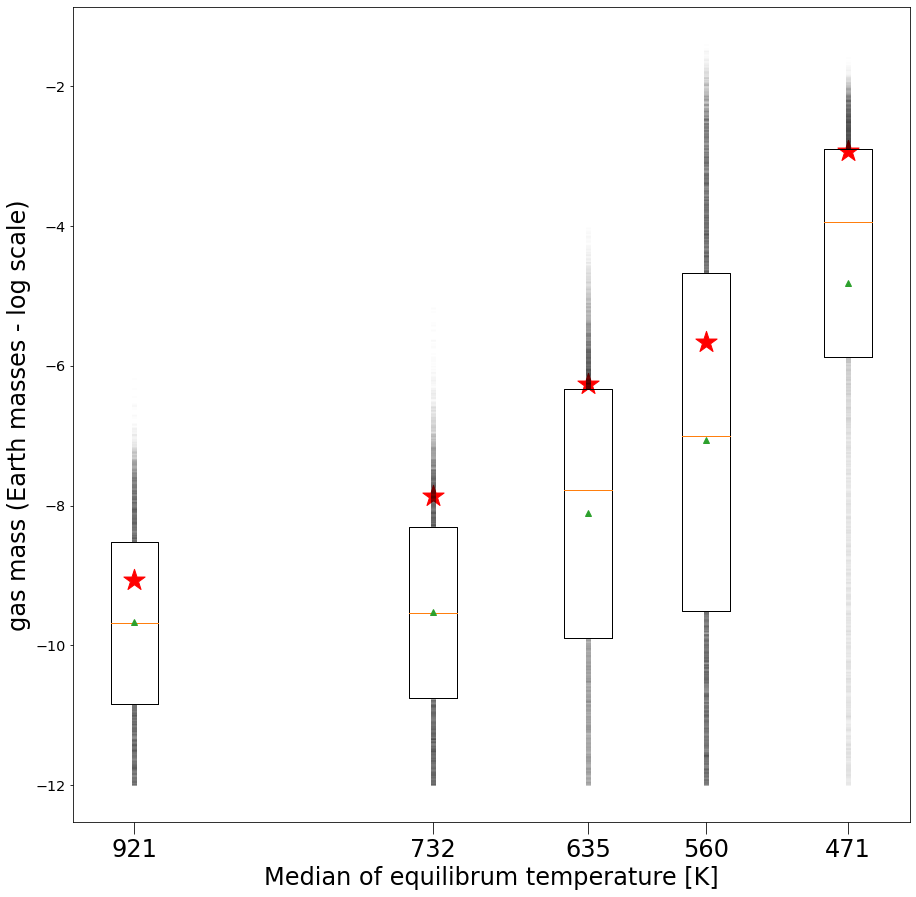}
 \quad
 \includegraphics[width=0.45 \hsize]{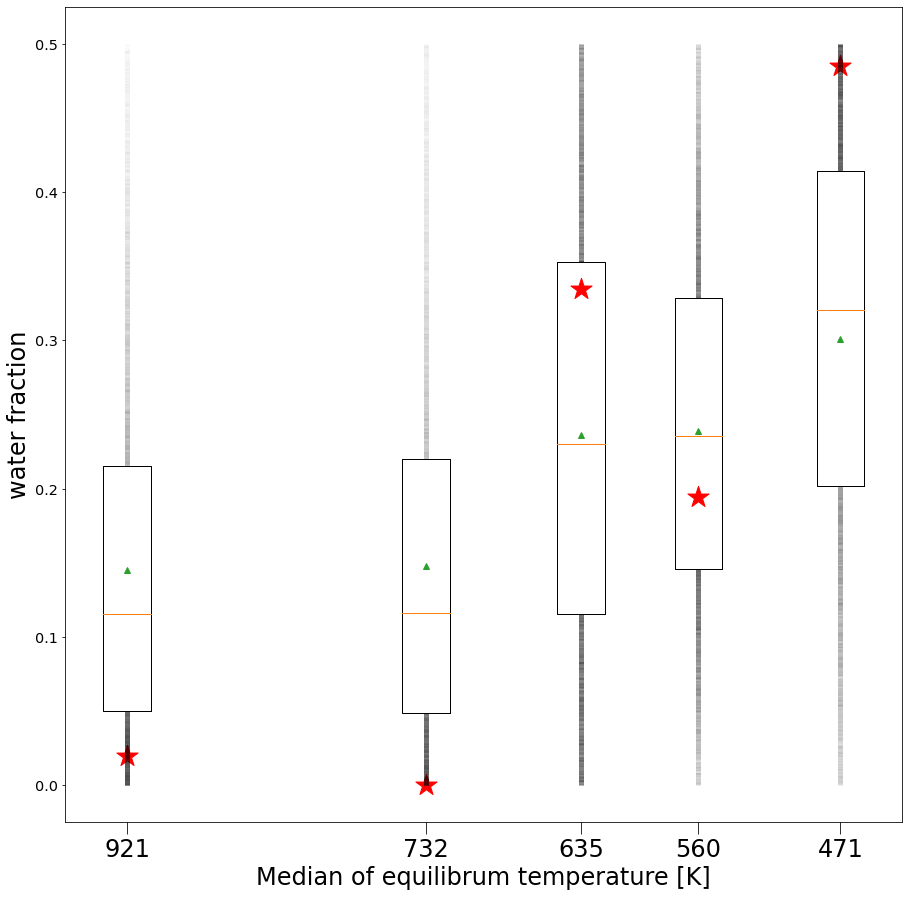}
\\
\includegraphics[width=0.45 \hsize]{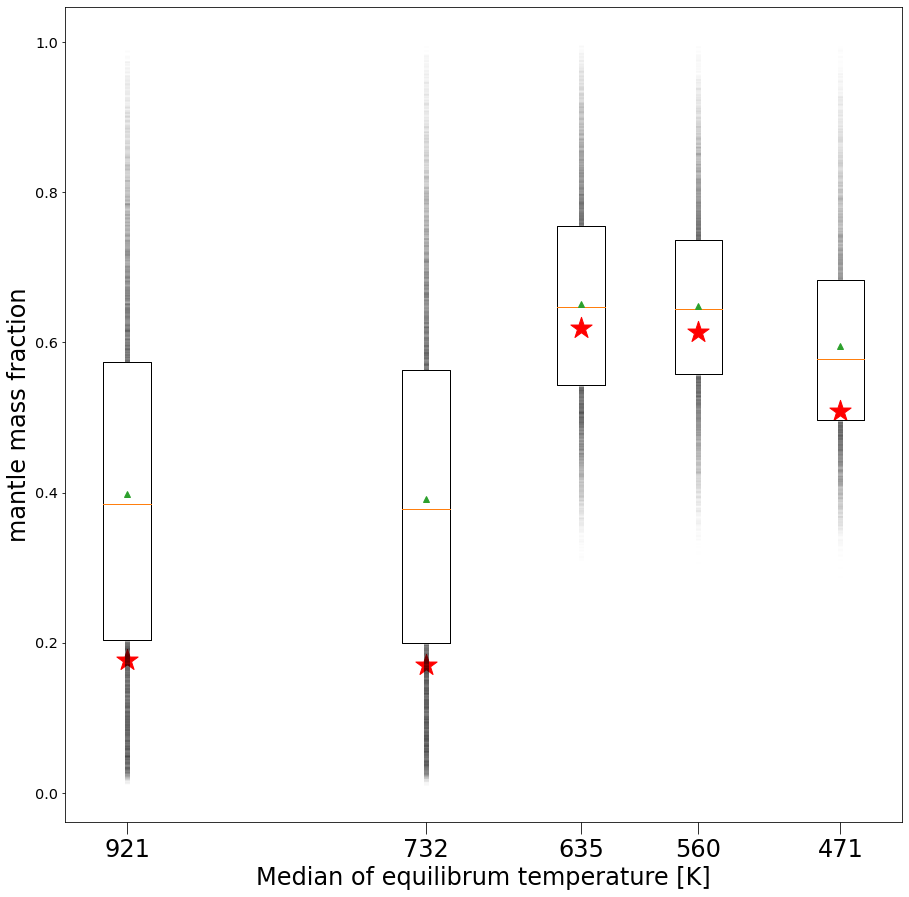}
\includegraphics[width=0.45 \hsize]{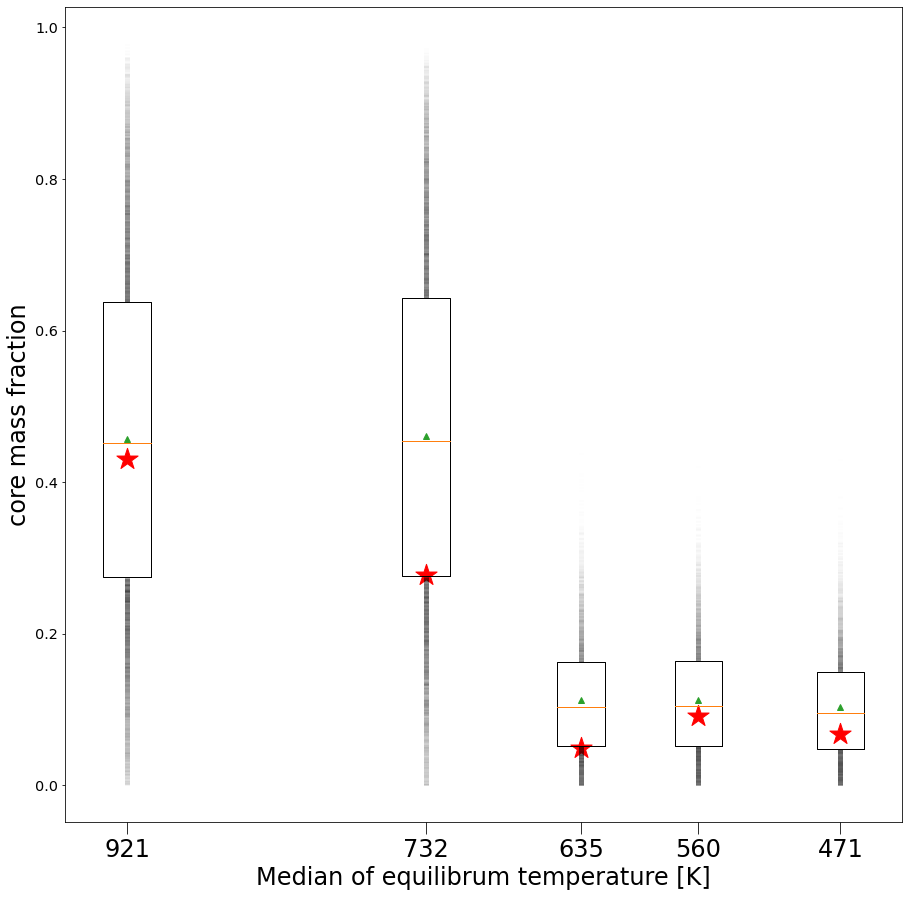}
  \caption{\label{comparatology}  Gas mass fraction (top left panel),  water fraction (top right panel), mantle mass fraction (bottom left panel), and core mass fraction (bottom right panel) for the five planets orbiting HD~23472.  The green triangle represents the distribution mean, the orange line represents the distribution median, the red star represents the distribution mode, and the box represents the 25 and 75 percentiles. The opacity of the black line (outside the boxes) is proportional to the posterior distribution (normalised to the mode).}
\end{figure*}

The gas fraction, water fraction, and core mass fraction for the system are shown in Figure \ref{comparatology} (see also Table \ref{comparatology_table}). These examples show a clear increase of the water mass fraction and  the gas mass fraction with decreasing planetary effective temperature. Our Bayesian analysis shows that the two inner planets have a small water mass fraction and a negligible gas mass fraction. The three outermost planets, particularly planets b and c, are likely to have much more water. Similarly, the three outermost planets could have a non-negligible mass of gas, up to a few percent of an Earth mass for the two outermost ones. The two innermost planets, on the other hand, are very likely to be devoid of gas. This could be explained by the loss of volatiles and gas due to irradiation on these planets. It could also be explained if the two inner planets formed inside the snow line and are dry, while the three outer planets formed beyond the snow line and are water worlds \citep{Venturini2020, Luque2022} . The two innermost planets have a large iron core (on the order of 30 to 60 \% in mass)  and have a Si/Fe and Mg/Fe ratio smaller than the stellar one, whereas the three outer planets have smaller cores (less than 20 \% in mass). This dichotomy in the structure of the planets (the two innermost ones versus the three outermost ones) is remarkable and puzzling in terms of identifying their formation processes (for more, see the next section).

\begin{table*}[h!]
\caption{Internal structure parameter for the five planets. 'Core' is the core mass fraction, 'Mantle' the mantle mass fraction, 'Water' the water mass fraction, 'Gas' the log (base 10) of the mass of gas (in Earth masses), 'Si/Fe' the Si/Fe bulk molar fraction and 'Mg/Fe/ the Mg/Fe bulk molar fraction. The uncertainties correspond to the 5 and 95 \% percentile, while the central value corresponds to the median. }
\begin{center}
\begin{tabular}{l  c c c c c c }
\hline
\hline
 Planet & Core & Mantle & Water & Gas  & Si/Fe  & Mg/Fe \\
\hline
HD~23472d & $0.45^{+0.38}_{-0.36}$ & $0.38^{+0.41}_{-0.33}$ & $0.12^{+0.27}_{-0.11}$ & $-9.68^{+2.10}_{-2.09}$ & $0.31^{+0.75}_{-0.28}$ & $0.28^{+0.77}_{-0.26}$ \\
HD~23472e & $0.45^{+0.39}_{-0.36}$ & $0.38^{+0.41}_{-0.32}$ & $0.12^{+0.28}_{-0.11}$ & $-9.54^{+2.30}_{-2.22}$ & $0.30^{+0.74}_{-0.27}$ & $0.28^{+0.77}_{-0.25}$ \\
HD~23472f  & $0.10^{+0.15}_{-0.09}$ & $0.65^{+0.24}_{-0.32}$ & $0.23^{+0.24}_{-0.21}$ & $-7.78^{+2.49}_{-3.80}$ & $1.26^{+0.37}_{-0.35}$ & $1.18^{+0.44}_{-0.43}$ \\
HD~23472b & $0.10^{+0.14}_{-0.09}$ & $0.64^{+0.21}_{-0.19}$ & $0.24^{+0.21}_{-0.19}$ & $-7.01^{+4.38}_{-4.50}$ & $1.26^{+0.37}_{-0.35}$ & $1.18^{+0.44}_{-0.43}$ \\
HD~23472c & $0.09^{+0.13}_{-0.09}$ & $0.58^{+0.26}_{-0.17}$ & $0.32^{+0.16}_{-0.26}$ & $-3.94^{+1.72}_{-6.68}$ & $1.26^{+0.37}_{-0.35}$ & $1.18^{+0.44}_{-0.43}$ \\
\hline
\hline
\end{tabular}
\end{center}
\label{comparatology_table}
\end{table*}

\subsection{Two likely super-Mercuries in HD~23472}

\begin{figure}[!htb]
 \centering
\includegraphics[width=0.98 \hsize]{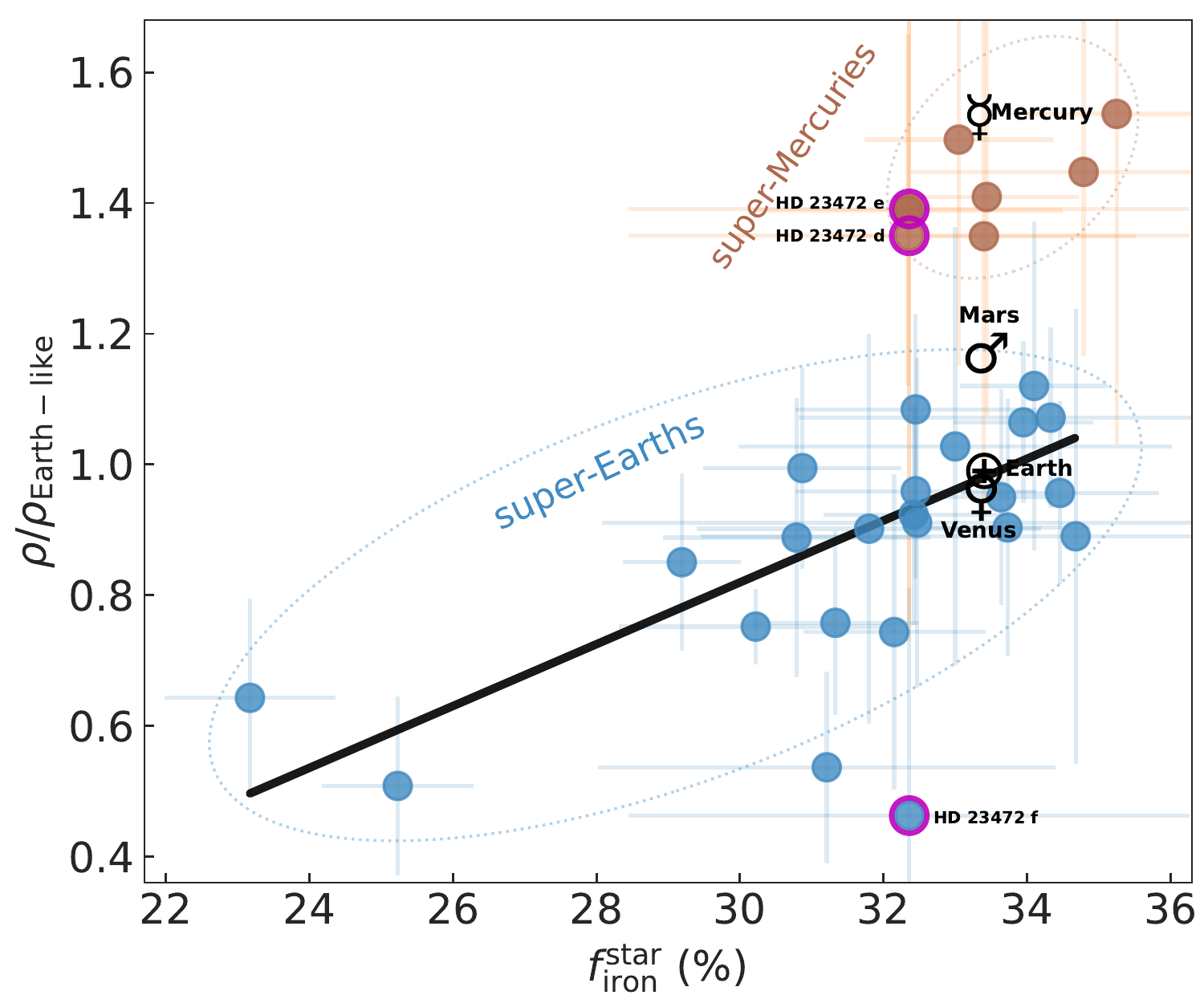}
  \caption{\label{vardan} Planetary normalised density of known rocky planets as a function of the estimated iron to silicate mass fraction of the protoplanetary disc. We also show the rocky Solar System planets represented by their symbols. For these, we also calculated the iron fraction from the abundances of the Sun.  HD~23472~d, HD~23472~e, and HD~23472~f planets which are below the radius gap are shown with a pink circle. We note that HD~137496 b is behind HD~23472 e in the figure since their positions almost overlap. The black solid line shows the correlation for the known exoplanets, excluding the potential super-Mercuries (brown circles) identified in \citet{Adibekyan2021}.}
\end{figure}

Recently, \citet{Adibekyan2021} reported that all super-Mercuries are formed in proto-planetary disks with enhanced iron abundance when compared to Mg and Si. Following \citet{Adibekyan2021} and using the stellar abundances, we estimated the iron-to-silicate-mass fraction of the proto-planetary disc of HD~23472. We found that it is high (32.3 +/- 3.9\%), despite the star being relatively metal-poor. We noticed that this occurs because the abundances of Si and Mg in the star are similar to the abundance of Fe. This is uncommon since most stars with low iron content are slightly enhanced in Si and Mg relative to iron due to galactic chemical evolution \citep{Adibekyan2012} . We also derived the planetary density normalised to the density of an Earth-like composition to check whether the planets of HD~23472 follow the correlation reported by  \citet{Adibekyan2021} between iron mass fraction of the star and the normalised density of super-Earths. The normalisation accounts for the increase in density with mass for planets with the same composition. Figure~\ref{vardan} compares the scaled planet density with the iron mass fraction of the star for planets HD~23472~d, HD~23472~e,  and HD~23472~f within the context of known small planets relation proposed by \citet{Adibekyan2021}. Planets b and c have significant water- or gas-rich envelopes (or a combination of both) and are above the radius gap; hence, they are not expected to follow the correlation. Planet f is below the radius gap but it might not follow the correlation if it has a significant gas rich envelope (see Fig.~\ref{comparatology}). 
The two inner planets are in the region of the super-Mercuries, which have higher densities than what would be expected from the host star composition. However, within the relative large mass uncertainties, they are also compatible with the Super-Earth population.

A better level of precision for the planetary mass is needed to confirm that both planets are super-Mercuries. If the presence of two super-Mercuries in HD~23472 is confirmed, it would make this an excellent test-bed for theories of super-Mercuries formation and evolution, and it may hold the key to solving their mystery. In the context of the giant impact theory, it has been shown that forming super-Mercuries would require a series of strong giant impacts \citep{Scora2020}. This implies that the formation of super-Mercuries via giant impact is rare, and the presence of two in the same system is extremely unlikely. The gap in planetary density between super-Mercuries and super-Earths reported by  \citet{Adibekyan2021} also calls into question the hypothesis of giant impact as its stochastic nature predicts a continuous distribution of densities. 
 
 The theory of mantle evaporation was proposed to explain the high density of Mercury, whose high dayside temperature would be sufficient to cause mantle evaporation into an atmosphere of silicate vapour \citep{Cameron1985, Perez-Becker2013}. However, a very high rate of evaporation is required, which is not supported by theory. Furthermore, this mechanism should be applicable to all super-Earths with similar equilibrium temperatures, which has not been observed \citep{Adibekyan2021}. 
 
 Photophoresis, that is, the depletion of silicates at the inner-edge of the proto-planetary disc, could also be responsible for the formation of super-Mercuries. Other mechanisms that also change the conditions of the inner disc, such as, rocklines \citep{Aguichine2020}, magnetic erosion \citep{Hubbard2014}, and magnetic boost \citep{Kruss2018}, have also been proposed to form super-Mercuries and might also be able two form two super-Mercuries in the same system.
  
 Finally, another possibility is that super-Mercuries are planetary cores that have 'recently' lost their envelopes due to a significant mass-loss event and are still compressed \citep{Mocquet2014}.  Characterising these cores would provide us with unique information about the planet's interiors.

\section{Conclusions}
\label{conclusion}
We obtained RV observations with the ESPRESSO spectrograph mounted in the VLT to measure the masses of the five planets around HD~23472. Combining our new observations with previous PFS RV observations and TESS photometry, we estimated the composition of the five planets in the system. We found slightly smaller masses for the two exoplanets that were previously confirmed, namely, planets b and c. We also constrained the mass of the other three smaller inner planets in the system. The two outermost planets are approximately twice the size of the Earth, and the mass of planet b, $M_b =8.42_{-0.84}^{+0.83}\,$ \MEarth,\ is more than twice the mass of planet c,  $M_c = 3.37_{-0.87}^{+0.92}\,$  \MEarth\ ; hence, it is much denser and has a lower gas and water content than planet c. The middle planet (planet f) is slightly larger than the Earth and the lightest of the planets, weighing approximately half as much as the Earth. The semi-amplitude of the RV signature of this planet was detected at $1.9\sigma$ (55\% relative precision) with an upper limit on the mass of 1.5  \MEarth\  at 95\% confidence level. Hence, more observations are required to confirm its low density and large water and gas mass fraction. The two inner planets are both smaller and lighter than Earth. The semi-amplitude of the RV signature of the two inner planets was detected at higher significance ($2.7\sigma$) but more observations are also advisable to confirm their derived high density. We show that their high density and properties match those of super-Mercuries previously discovered. Using Bayesian internal structure model we find that the two inner planets have much higher iron core than the outer planets and they have a Si/Fe and Mg/Fe ratio smaller than the stellar one consistent with super-Mercuries composition. Further RV observations to improve the precision of the planetary masses would greatly improve our understanding of this unique system. If they confirm that the composition of the two inner planets, this would be the first time two super-Mercuries have been discovered in the same system, making it a golden target for further characterisation.

\begin{figure}[!htb]
 \centering
\includegraphics[width=0.98 \hsize]{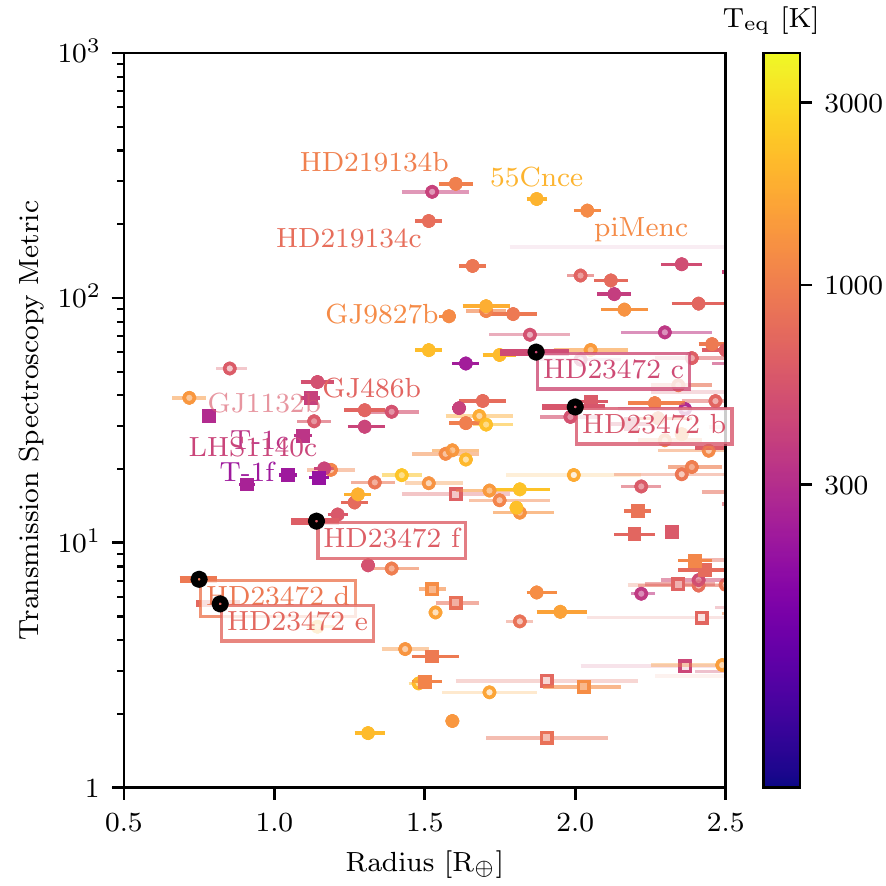}
  \caption[]{\label{jwst} Transmission spectrum metric as a function of planetary radius for the five planets in the HD~23472 and the well characterised small planet population. We show all planets from the exoplanet archive with precision in mass and radius better than 50\%\. The planets with mass derived by RVs are shown as circles and the planets with mass derived by TTVs are shown as squares. The color of the points indicates the planet effective temperature. The planets in the HD~23472 are clearly marked. This graph was created using the mass radius diagram code\footnotemark}
\end{figure}

 Atmosphere observations may shed light on the formation of the two super-Mercuries as well as the system architecture. We computed the transmission spectrum metric (TSM, \citealt{Kempton2018}) to access the observability with the James Webb space telescope. Planets b, c, d, e, and f have TSMs of 36, 59, 7, 5, and 14, respectively. It is only planet f that has a TSM greater than the threshold proposed by \citet{Kempton2018}. The TSM values for the five planets of HD~23472 system are shown in Figure~\ref{jwst}, set in the context of known and well-characterised exoplanets. Planets b and c are good targets for transmission spectroscopy due to their low effective temperature  $T_{\textrm{eq}} < 550\,$K.  As expected, the two likely super-Mercuries are difficult targets for transmission spectroscopy, but they may be better targets for emission spectroscopy using, for example, the future spectrograph ANDES@ELT. Furthermore, HD~23472 is the brightest star with super-Mercuries (H mag = 7.3), making this the best system for studying a potential atmosphere around Super-Mercuries. For an Earth-mass planet orbiting HD~23472,  the optimistic limits of the habitable zone correspond to periods between 98 days and 355 days. If the near-resonant chain continues to longer orbital periods and the system contains two longer period planets (g and h), planet h may be on the edge of the optimistic habitable zone. As a result, HD23472 is shown to be an excellent candidate for the search for habitable Earths. We also encourage monitoring studies of the transit times of the two outermost planets to assess whether they are in MMR.

\footnotetext{The code is available online at \url{https://github.com/odemangeon/mass-radius_diagram}}.

\begin{acknowledgements}
The authors acknowledge the ESPRESSO project team for its effort and dedication in building the ESPRESSO instrument.

This research has made use of the NASA Exoplanet Archive, which is operated by the California Institute of Technology, under contract with the National Aeronautics and Space Administration under the Exoplanet Exploration Program.

This work has made use of data from the European Space Agency (ESA) mission (\href{https://www.cosmos.esa.int/gaia}{\it Gaia}), processed by the {\it Gaia} Data Processing and Analysis Consortium (\href{https://www.cosmos.esa.int/web/gaia/dpac/consortium}{\textsc{dpac}}).
Funding for the \textsc{dpac} has been provided by national institutions, in particular the institutions participating in the {\it Gaia} Multilateral Agreement.

This work was supported by FCT - Funda\c{c}\~{a}o para a Ci\^{e}ncia - through national funds and by FEDER through COMPETE2020 - Programa Operacional Competitividade e Internacionalização by these grants: UID/FIS/04434/2019; UIDB/04434/2020; UIDP/04434/2020; PTDC/FIS-AST/32113/2017 \& POCI-01-0145-FEDER-032113; PTDC/FIS-AST/28953/2017 \& POCI-01-0145-FEDER-028953; PTDC/FIS-AST/28987/2017 \& POCI-01-0145-FEDER-028987; PTDC/FIS-AST/30389/2017 \& POCI-01-0145-FEDER-030389. CJAPM acknowledges FCT and POCH/FSE (EC) support through Investigador FCT Contract 2021.01214.CEECIND/CP1658/CT0001.

O.D.S.D.~is supported in the form of work contract (DL 57/2016/CP1364/CT0004) funded by FCT.
JIGH, RR, CAP and ASM acknowledge financial support from the Spanish Ministry of Science and Innovation (MICIN) project PID2020-117493GB-I00. ASM, JIGH and RR also acknowledge financial support from the Government of the Canary Islands project ProID2020010129.

This work has been carried out with the support of the framework of the National Centre of Competence in Research PlanetS supported by the Swiss National Science Foundation (SNSF). The authors acknowledge the financial support of the SNSF and in particular YA and JH acknowledge the SNSF for supporting research through the grant 200020\_19203.




A.S. acknowledges support from the Italian Space Agency (ASI) under contract 2018-24-HH.0. The financial contribution from the agreement ASI-INAF n.2018-16-HH.0 is gratefully acknowledged.

This project has received funding from the European Research Council (ERC) under the European Union’s Horizon 2020 research and innovation programme (project {\sc Four Aces}; grant agreement No 724427).  It has also been carried out in the frame of the National Centre for Competence in Research PlanetS supported by the Swiss National Science Foundation (SNSF). DE acknowledges financial support from the Swiss National Science Foundation for project 200021\_200726.
FPE and CLO would like to acknowledge the Swiss National Science Foundation (SNSF) for supporting research with ESPRESSO through the SNSF grants nr. 140649, 152721, 166227 and 184618. The ESPRESSO Instrument Project was partially funded through SNSF’s FLARE Programme for large infrastructures.

J.L-B. acknowledges financial support received from "la Caixa" Foundation (ID 100010434) and from the European Unions Horizon 2020 research and innovation programme under the Marie Slodowska-Curie grant agreement No 847648, with fellowship code LCF/BQ/PI20/11760023. This research has also been partly funded by the Spanish State Research Agency (AEI) Projects No.PID2019-107061GB-C61 and No. MDM-2017-0737 Unidad de Excelencia "Mar\'ia de Maeztu"- Centro de Astrobiolog\'ia (INTA-CSIC).

  We acknowledge the use of public TESS data from pipelines at the TESS Science Office and at the TESS Science Processing Operations Center.

Resources supporting this work were provided by the NASA High-End Computing (HEC) Program through the NASA Advanced Supercomputing (NAS) Division at Ames Research Center for the production of the SPOC data products.

We thank Lisa Kaltenegger for useful discussions.

\end{acknowledgements}

\bibliographystyle{aa} 

\bibliography{susana}

\begin{appendix}

\section{Additional tables and figures for our best-fit model}

The best fit parameters of our simultaneous RV (ESPRESSO + PFS data) and light curve (TESS) model for the 5 planet system HD23472 are shown in Table~\ref{syspar}. To demonstrate the precision of the new observations with ESPRESSO, we also show the best fit most relevant parameters for our model, using only the new ESPRESSO data and the light curve in Table~\ref{sysparesp}. As stated in Section~\ref{performance}, the results (including or not including the PFS data) are within $1 sigma $ of each other. The precision of the RV+amplitude of all planets improves as expected when the 64 RV observations of PFS are added. The priors for the model are given in Table~\ref{priors}.

\begin{figure*} 
\centering 
\includegraphics[width=1.9\columnwidth]{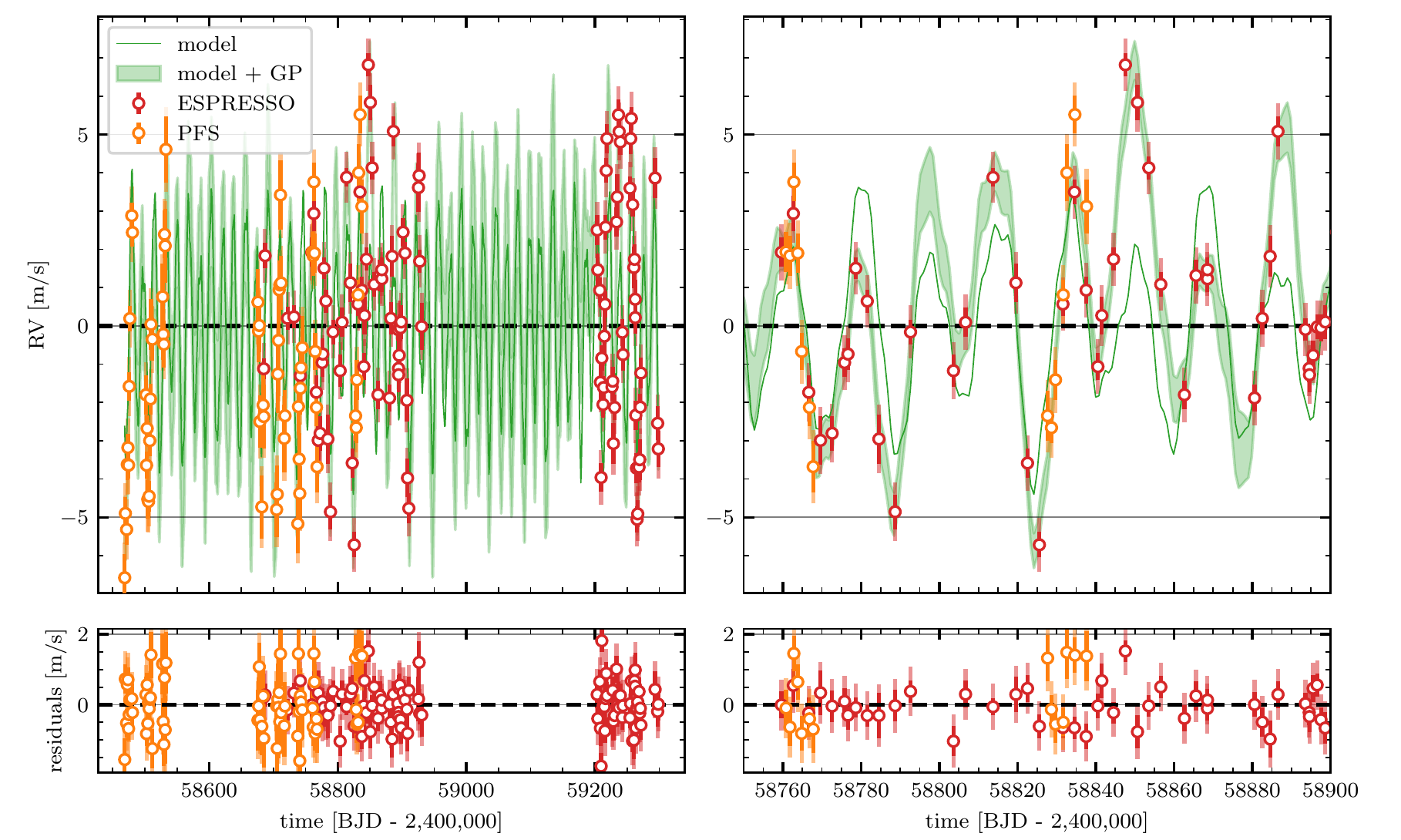}
\caption{Time series of the RVs of ESPRESSO (in red) and PFS (in orange) together with the best fit model (in green) with five planets and GP to account for stellar activity (left panel). We show the $1\,\sigma$ uncertainties from the GP model in shaded green. The systemic velocity and the offset between the two instruments was corrected. The residuals of the best model are shown in the bottom panel. Zoom of the left panel is shown on the right, displaying the middle season of observations when the ESPRESSO and the PFS observations overlap. \label{RV_modelzoom}} 
\end{figure*}

Figure~\ref{RV_modelzoom} depicts the RV time series, as well as the best-fit five-planet model and the GP activity model. The relative high RV variability due to stellar activity is  clearly evident. We also show a zoom of the plot during the overlap of ESPRESSO and PFS observations, demonstrating good agreement between the two instruments.

\begin{figure} 
\centering 
\includegraphics[width=0.95\columnwidth, trim={0 0 0.7cm 0} , clip  ]{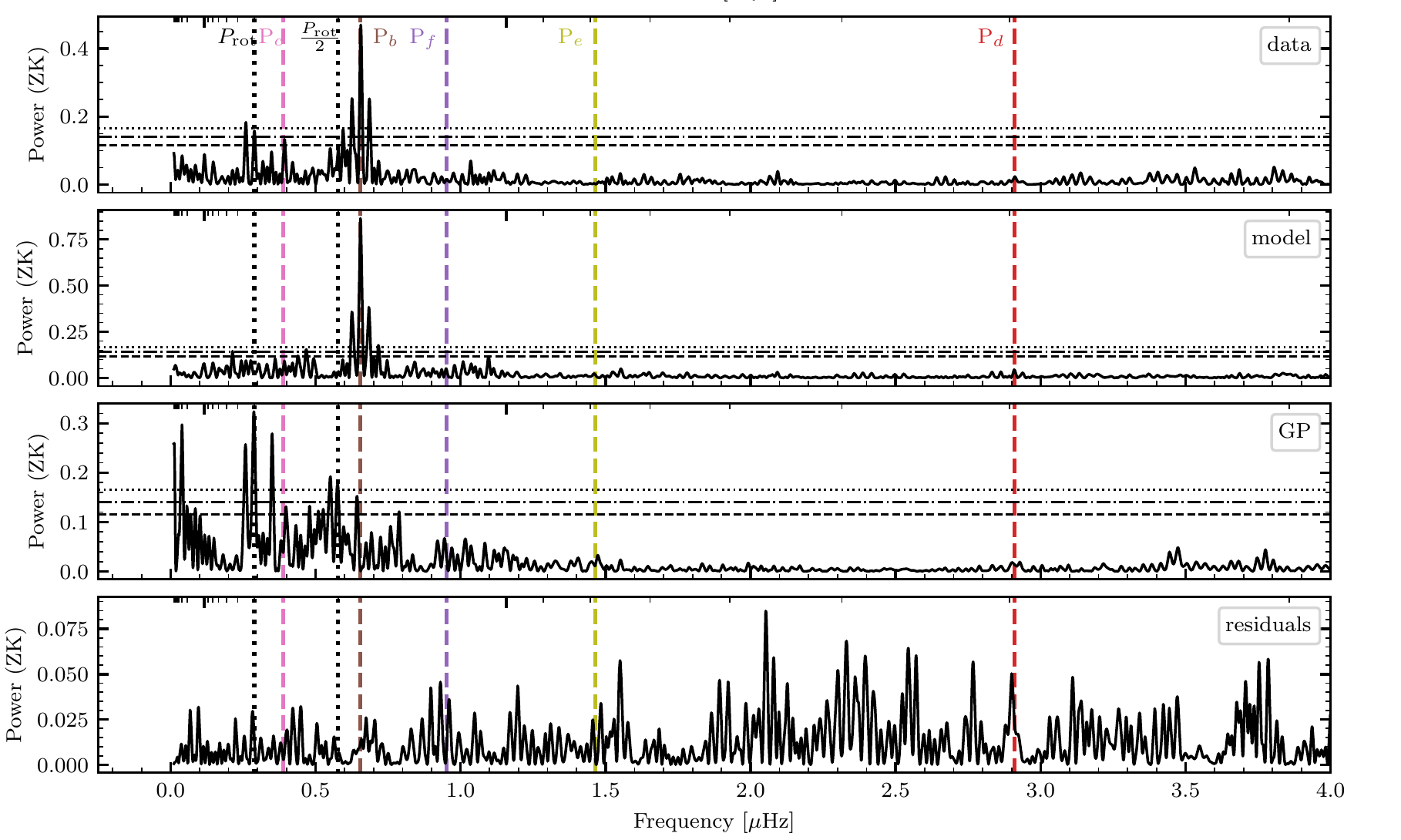}
\caption{GLS periodogram of the RV data (top panel), the five Keplerian model (second panel), the GP model (third panel), and the residuals (fourth panel). The horizontal lines show the 10\% (dashed line), 1\% (dot-dashed line), and 0.1\% (dotted line) FAP levels calculated following \citet{Zechmeister2009}. The coloured dotted vertical lines show the position of the known transiting planets while the black dotted vertical lines show the position of the estimated rotation period of the star and its first harmonic. \label{GLS_model}} 
\end{figure}

In Figure~\ref{GLS_model}, we compare the GLS of the RV data with the GLS of the five Keplerian model, the GP model, and the residuals. With the exception of planet c, which is modelled by the Keplerian model, the GP aptly models all the periodicities longer than half the rotation period of the star ($P_\mathrm{rot} \sim 40\,$days). The residuals show low power periodicities at shorter time scales, which could be due to short timescale stellar variability such as granulation.

\begin{figure} 
\centering 
\includegraphics[width=0.95\columnwidth]{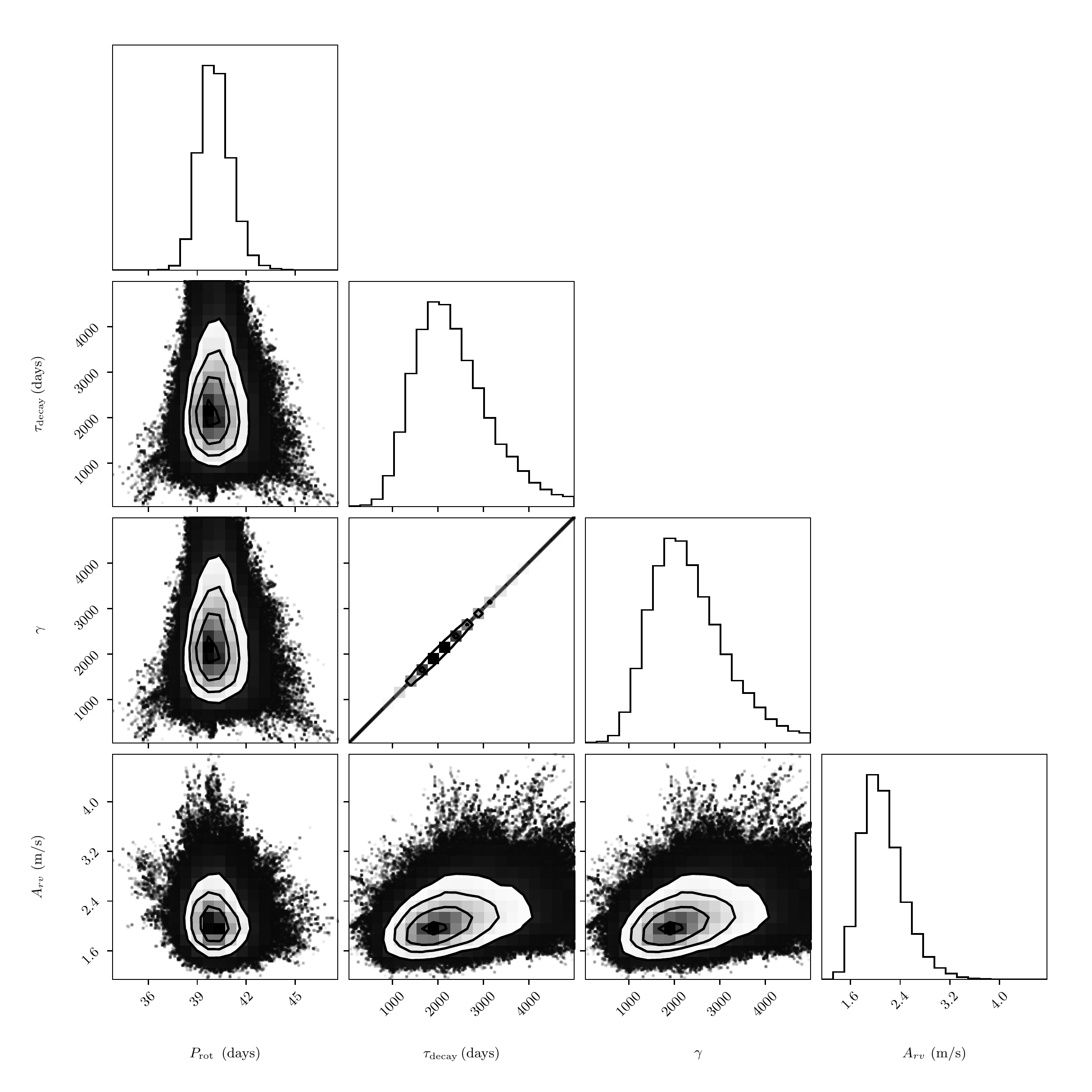}
\caption{Corner plot of the hyper-parameters of the GP that models the RV activity signal. All the hyper-parameters: the period of the activity signal ($P_\mathrm{rot}$), the decay timescale ($\tau_{\mathrm{decay}}$), the periodic coherence scale ( $\gamma$ )  and the amplitude of the activity signal ($A_{\rv}$)  are well constrained.  \label{cornerhyper}} 
\end{figure}

Figure~\ref{cornerhyper} shows the corner plots of the GP model's hyper-parameters. The posteriors of the GP hyper-parameter demonstrate that the RV aptly data constrain hyper-parameters.

\onecolumn

\begin{symbolfootnotes}
\begin{raggedleft}
\begin{longtable}{p{0.25\textwidth}P{0.25\textwidth}P{0.25\textwidth}P{0.25\textwidth}}
\caption{ \textbf{Parameter estimates for the planetary system HD~23472} \label{syspar} } \\ %
\hline
\endfirsthead
{\tablename\ \thetable\ -- \textit{Continued from previous page}} \\
\hline
\endhead
\hline \textit{Continued on next page}\\
\endfoot
\hline\\
\endlastfoot
\textit{Planetary parameters} \\
\hline
  \\[-3pt]
& \textit{Planet b} & \textit{Planet c} \\ 
  \\[-5pt]
$M_p$ [\MEarth]                                         &$8.32_{-0.79}^{+0.78}$  & $3.41_{-0.81}^{+0.88}$\\
$R_p$ [\REarth]                                         & $2.00_{-0.10}^{+0.11}$& $1.87_{-0.11}^{+0.12}$\\
$\rho_p$ [$\mathrm{g.cm^{-3}}$]               & $6.15_{-1.0}^{+1.2}$ & $3.1_{-0.84}^{+1.0}$\\
$T_{\textrm{eq}}$ [K]                                   & $543\pm 18$&$467_{-18}^{+19}$\\
${P}^{\ \bullet}$\ [days]                               &  $17.667087\pm0.000042$&  $29.79749_{-0.00014}^{+0.00013}$\\
${t_{\textrm{ic}}}^{\bullet}$\ [BJD$_{\mathrm{TDB}}$ - 2\,457\,000]   &  $1360.6641\pm 0.0012$ &  $1370.1037_{-0.0018}^{+0.0019}$\\
$a$ [AU]                                              &$0.1162\pm0.0018$ &$0.1646 \pm 0.0024$\\
$e$                                                     &$0.072_{-0.040}^{+0.039}$ &$0.063_{-0.043}^{+0.054}$\\
$\omega_*$ [$^\circ$]                                  &  $114_{-210}^{+37}$&  $2_{-128}^{+116}$\\
${K}^{\bullet}$ [\ms]                                   & $2.68_{-0.25}^{+0.24}$& $0.92\pm 0.23 $\\
$i_p$ [$\deg$]                                                       &   $88.93\pm0.16$&    $89.095_{-0.073}^{+0.089}$\\
${R_p / R_*}^{\bullet}$                                             & $0.02597_{-0.00083}^{+0.00076}$& $0.0243_{-0.0010}^{+0.0011}$\\
$a / R_*$                                               &  $37.1_{-1.9}^{+2.1}$& $50.2_{-3.1}^{+3.5}$\\
$b$                                                     &$0.690_{-0.090}^{+0.054}$&$0.785_{-0.054}^{+0.043}$\\
$D14$ [h]                                               &  $2.67_{-0.17}^{+0.11}$&   $2.94\pm 0.33$\\
$D23$ [h]                                                          &      $2.42_{-0.17}^{+0.11}$&      $2.59_{-0.28}^{+0.30}$\\
$F_{i}$ [$F_{i, \oplus}$]                                       &$16.0_{-1.6}^{+1.8}$&$7.96_{-0.78}^{+0.87}$\\
$H$ [km]                                                & $101_{-15}^{+18}$ & $186_{-43}^{+68}$\\                                           
  \\[-2pt]
 & \textit{Planet d} &  \textit{Planet e}  &  \textit{Planet f} \\
   \\[-5pt]
$M_p$ [\MEarth]                                        & $0.55_{-0.20}^{+       0.21}$ &$0.72_{-0.27}^{+0.28}$&$0.77_{-0.40}^{+0.44}$\\
$R_p$ [\REarth]                                         & $0.750_{-0.057}^{+    0.067}$& $0.818_{-0.065}^{+      0.080}$& $1.137_{-0.077}^{+     0.084}$\\
$\rho_p$ [$\mathrm{g.cm^{-3}}$]               & $7.5_{-3.1}^{+3.9}$& $7.5_{-3.0}^{+3.9}$& $3.0_{-1.6}^{+2.0}$\\
$T_{\textrm{eq}}$ [K]                                   & $909_{-32}^{+35}$ & $723_{-25}^{+28}$& $630_{-23}^{+27}$\\
${P}^{\ \bullet}$\ [days]                               & $3.97664_{-0.000044}^{+0.000030}$ &  $7.90754\pm 0.00011 $&  $12.1621839_{0.000099}^{+0.00012}$\\
${t_{\textrm{ic}}}^{\bullet}$\ [BJD$_{\mathrm{TDB}}$ - 2\,457\,000]   &  $1357.8398_{-0.0062}^{+0.0059}$&  $1354.5659 \pm 0.0066$&  $1360.0754_{-0.0078}^{+0.0049}$\\
$a$ [AU]                                              & $0.04298_{-0.00065}^{+0.00063}$&$0.0680  \pm 0.0010$&$0.0906\pm0.0014$\\
$e$                                                     & $0.070_{-0.047}^{+0.050}$&$0.070_{-0.047}^{+0.052}$&$0.070_{-0.051}^{+0.048}$\\
$\omega_*$ [$^\circ$]                                  & $19_{-117}^{+99}$&  $46_{-179}^{+100}$& $4_{-133}^{+97}$\\
${K}^{\bullet}$ [\ms]                                   & $0.29\pm 0.11 $& $0.30_{-0.11}^{+0.12}$ & $0.29_{-0.11}^{+0.11}$\\
$i_p$ [$\deg$]                                                          & $87.95_{-0.87}^{+1.2}$&    $88.63_{-0.56}^{+0.80}$&   $88.81_{-0.32}^{+0.58}$\\
${R_p / R_*}^{\bullet}$                                             & $0.009752_{-0.00072}^{+0.00077}$& $0.01065_{-0.00076}^{+0.00088}$&$0.01478_{-0.00077}^{+0.00087}$\\
$a / R_*$                                               & $13.29_{-0.78}^{+0.79}$&  $21.0_{-1.2}^{+1.3}$&  $28.0_{-2.0}^{+1.9}$\\
$b$                                                     & $0.47_{-0.28}^{+0.16}$ &$0.50_{-0.29}^{+0.18}$&$0.57_{-0.26}^{+0.10}$\\
$D14$ [h]                                               & $1.99_{-0.20}^{+0.21}$ &   $2.48 \pm0.30$& $2.82_{-0.14}^{+0.22}$\\
$D23$ [h]                                                                   & $1.94_{-0.20}^{+0.21}$ &     $2.41\pm0.30$ &        $2.70_{-0.15}^{+0.21}$\\
$F_{i}$ [$F_{i, \oplus}$]                                       & $117_{-11}^{+13}$& $46.7_{-4.6}^{+5.1}$&$26.3_{-2.6}^{+2.9}$\\
$H$ [km]                                                & $363_{-124}^{+282}$ &$265_{-82}^{+164}$& $412_{-210}^{+713}$\\ 
 \\[-3pt] 
\textit{Stellar parameters}\\
\hline\hline
${v0}^{\bullet}$ [\kms]                                              & $34.55274 \pm 0.00055$\\
${\rho_*}^{\bullet}$ [$\rho_\sun$]             & $1.92\pm 0.18$  & \\
${A_{\rv}}^{\bullet}$ [\ms]                                         & $2.079_{-0.28}^{+0.37}$\\
${P_{\mathrm{rot}}}^{\bullet}$ [days]                         & $40.1_{-0.87}^{+1.0}$\\
${\tau_{\mathrm{decay}}}^{\bullet}$ [days]               &  $2218_{-684}^{+961}$\\
${\gamma}^{\bullet}$                                                & $0.385_{-0.060}^{+0.071}$\\
$u_{1,\tess}^{\bullet}$                                         & $0.496 \pm 0.051$\\
$u_{2,\tess}^{\bullet}$                                         & $0.210 \pm 0.052$\\
\\[-5pt]
\multicolumn{3}{l}{\textit{Parameters of instruments}} \\
\hline\hline
$\Delta\mathrm{RV}_{\mathrm{PFS/ESPRESSO}}^{\bullet}$  [\kms]       &  $-34.55123 \pm 0.00030$\\
$\sigma_{\rv, \mathrm{ESPRESSO}}^{\bullet}$ [\ms]                             & $0.617_{-0.088}^{+0.098}$   \\ 
$\sigma_{\rv, \mathrm{PFS}}^{\bullet}$ [\ms]                             & $0.678_{-0.14}^{+0.15}$   \\ 
$\sigma_{\tess}^{\bullet}$ [ppm]                                              &  $31_{-21}^{+24}$ &  \\ 
\hline
\end{longtable}
\tablefoot{\\
$^{\bullet}$ indicates that the parameter is a main or jumping parameter for the \textsc{mcmc} explorations\\
}
\end{raggedleft}
\end{symbolfootnotes}

\begin{symbolfootnotes}
\begin{raggedleft}
\begin{longtable}{p{0.15\textwidth}P{0.15\textwidth}P{0.15\textwidth}P{0.15\textwidth}P{0.15\textwidth}P{0.15\textwidth}}
\caption{ \textbf{Parameter estimates for the planetary system HD~23472 - ESPRESSO data only \label{sysparesp} }} \\ %
\hline
\endfirsthead
{\tablename\ \thetable\ -- \textit{Continued from previous page}} \\
\hline
\endhead
\hline \textit{Continued on next page}\\
\endfoot
\hline\\
\endlastfoot
\multicolumn{3}{l}{\textit{Planetary parameters}} \\
\hline
  \\[-3pt]
 & \textit{Planet b}& \textit{Planet c} & \textit{Planet d}& \textit{Planet e} & \textit{Planet f} \\
$M_p$ [\MEarth]                                         &$8.54_{-1.04}^{+0.99}$\  & $2.77_{-0.96}^{+0.96}$ & $0.45_{-0.25}^{+    0.28}$&$0.51_{-0.29}^{+0.33}$ &$0.89_{-0.46}^{+       0.54}$ \\
$R_p$ [\REarth]                                         & $2.06_{-0.10}^{+0.11}$ & $2.06_{-0.10}^{+0.11}$& $0.749_{-0.058}^{+    0.067}$  & $0.820_{-0.066}^{+   0.076}$  &$1.135_{-0.081}^{+    0.089}$\\
$\rho_p$ [$\mathrm{g.cm^{-3}}$]               & $5.75_{-1.0}^{+1.1}$   & $2.58_{-0.96}^{+1.1}$& $6.3_{-3.4}^{+3.8}$  & $5.2_{-3.0}^{+4.1}$  & $3.5_{-1.8}^{+2.3}$\\ 
$e$                                                     &$0.052_{-0.033}^{+0.039}$&$0.099_{-0.068}^{+0.066}$& $0.086_{-0.058}^{+0.068}$&$0.092_{-0.063}^{+0.069}$  &$0.079_{-0.053}^{+0.063}$ \\
$\omega_*$ [$^\circ$]                                  &  $-94_{-60}^{+246}$ &  $38_{-128}^{+78}$ & $35_{-151}^{+105}$     &  $43_{-155}^{+102}$   & $11_{-135}^{+120}$\\
${K}^{\bullet}$ [\ms]                                   & $2.75_{-0.33}^{+0.31}$ & $0.75\pm 0.26 $  & $0.24_{-0.14}^{+0.15}$ & $0.216_{-0.124}^{+        0.140}$ & $0.32_{-0.17}^{+0.20}$ \\
\\[-3pt]
\hline
\multicolumn{3}{l}{\textit{Stellar parameters}}\\
\hline
${v0}^{\bullet}$ [\kms]                                              & $34.55303 \pm 0.00062$\\
${A_{\rv}}^{\bullet}$ [\ms]                                         & $2.079_{-0.28}^{+0.37}$\\
${P_{\mathrm{rot}}}^{\bullet}$ [\ms]                         & $40.5_{-1.1}^{+1.4}$\\
${\tau_{\mathrm{decay}}}^{\bullet}$ [days]               &  $2158_{-682}^{+924}$\\
${\gamma}^{\bullet}$                                                & $0.373_{-0.053}^{+0.062}$\\
\hline
\end{longtable}
\end{raggedleft}
\end{symbolfootnotes}

\begin{symbolfootnotes}
\begin{raggedleft}
\begin{longtable}{p{0.45\textwidth}P{0.45\textwidth}}
\caption{ \textbf{Priors of the global fit}\label{priors} } \\ %
\hline
\endfirsthead
{\tablename\ \thetable\ -- \textit{Continued from previous page}} \\
\hline
\endhead
\hline \textit{Continued on next page}\\
\endfoot
\hline\\
\endlastfoot
\textit{Planetary parameters} \\
\hline
  \\[-3pt]
 \textit{Common priors} \\
$K$ [\ms]                                   & $  \mathcal{U}(0, 33)$ \\
$R_p / R_*$                                & $  \mathcal{U}(0, 1)$ \\   
$b$                                                           &   $\mathcal{U}(0, 1)$ \\        
$e$                                                     & $ \mathcal{J}(1e-10, 0.15)$ \\
$\omega_*$ [$^\circ$]                         &         $ \mathcal{U}(-180, 180)$ \\
  \\[-3pt]
\textit{Planet b} \\ 
  $P$\ [days]                              &  $ \mathcal{N}( 17.66709, \num{4.e-4} )$ \\
 Mid-transit phase  ($T0= 58360.66359619588$ days) &    $ \mathcal{N}( 0, \num{6.22e-4})$ \\
  \\[-3pt]
 \textit{Planet c} \\ 
  $P$\ [days]                               &  $ \mathcal{N}( 29.7975 , 0.0016)$ \\
 Mid-transit phase  ($T0= 58370.103523632475$ days)  & $ \mathcal{N}( 0,  \num{6.37e-04})$ \\
  \\[-3pt]
  \textit{Planet d} \\
  $P$\ [days]                              &  $ \mathcal{N}( 3.9766 ,\num{3.4e-4})$ \\
 Mid-transit phase  ($T0= 58357.83648843389$ days) &   $ \mathcal{N}( 0,  0.015)$ \\                             
 \\[-3pt] 
  \textit{Planet e} \\
    $P$\ [days]                               &  $ \mathcal{N}( 7.9074 ,  0.0003)$ \\
 Mid-transit phase  ($T0= 58354.581189835175$ days) &    $ \mathcal{N}( 0, 0.0046)$ \\ 
 \\[-3pt]
 \textit{Planet f} \\
    $P$\ [days]                               &  $ \mathcal{N}( 12.16217 , \num{9.5e-4})$ \\
 Mid-transit phase  ($T0= 58360.07452141293$ days) &    $ \mathcal{N}( 0, 0.0066)$ \\ 
\\[-3pt]
\textit{Stellar parameters}\\
\hline\hline
${v0}$ [\kms]                                               &  $  \mathcal{U}(34.543230764, 34.559560714)$ \\
${\rho_*}$ [$\rho_\sun$]                  & $ \mathcal{N}(1.88,  0.18 )$ \\
${A_{\rv}}$ [\ms]                                          &  $  \mathcal{U}(0, 33)$ \\
${P_{\mathrm{rot}}}$ [days]                         &  $  \mathcal{U}(20, 100)$ \\
${\tau_{\mathrm{decay}}}$ [days]             &  $  \mathcal{U}(2.5, 5000)$ \\
${\gamma}$                                              &  $  \mathcal{U}(0.05, 5)$ \\
$u_{1,\tess}$                                         &$ \mathcal{N}( 0.472, 0.056 )$ \\
$u_{2,\tess}$                                         &$ \mathcal{N}( 0.186,0.055 )$ \\
\\[-5pt]
\multicolumn{2}{l}{\textit{Parameters of instruments}} \\
\hline\hline
$\Delta\mathrm{RV}_{\mathrm{PFS/ESPRESSO}}$  [\kms]       &  $  \mathcal{U}(-34.6, -34.4)$ \\
$\sigma_{\rv, \mathrm{ESPRESSO}}$ [\ms]                             &  $  \mathcal{U}(0, 2)$ \\
$\sigma_{\rv, \mathrm{PFS}}$ [\ms]                             &  $  \mathcal{U}(0, 3)$ \\
$\sigma_{\tess}$ [ppm]                                              &  $  \mathcal{U}(0, 400)$ \\
\hline
\end{longtable}
\tablefoot{\\
$\mathcal{U}(a;b)$ is a uniform distribution between $a$ and $b$; $\mathcal{J}(a;b)$ is a Jeffreys distribution between $a$ and $b$; $\mathcal{N}(a;b)$ is a
normal distribution with mean $a$ and standard deviation $b$. 
}
\end{raggedleft}
\end{symbolfootnotes}

\twocolumn

\section{Further periodicity analysis of the RVs}

To search for additional RV candidates, we analysed the ESPRESSO and PFS RV data with the $\ell_1$ periodogram \citep{Hara2017}. This tool is based on a sparse recovery technique called the basis pursuit algorithm \citep{Chen1998}. The $\ell_1$  periodogram takes in a frequency grid and an assumed covariance matrix of the noise as inputs\footnote{The code is available online at \url{https://github.com/nathanchara/l1periodogram}.}. It aims to find a representation of the RV time series as a sum of a small number of sinusoids whose frequencies are in the input grid. It produces as output a figure which has a similar aspect as a GLS periodogram, but with fewer peaks due to aliasing. FAPs can be calculated for each peak, whose interpretation is equivalent to that of common periodograms. 

To determine the influence of the noise model, we followed \citep{Hara2020} and considered a grid of covariance models, then we ranked the alternatives with a cross-validation process. We defined the covariance matrix, $V,$ so that its element at index $k,l$ is as follows:
\begin{equation}
\begin{split}
V_{k,\,l}  =  \delta_{k,l} (\sigma_{RV}^2 + \sigma_{W}^2 + \sigma_{C}^2 ) +   \sigma_{R}^2 \e^{[-\frac{(t_k-t_l)^2}{2\tau_R^2}]} + \nonumber \\
\sigma_{QP}^2\e^{\Big[-\frac{(t_k-t_l)^2}{2\tau_{act}^2} -\frac{1}{2} \sin^2\left(  \frac{ \pi (t_k - t_l)}{P_{act} } \right)\Big] }
\end{split}
\label{eq:kernel_l1},
\end{equation}

where $\sigma_{RV}$ is the nominal measurement uncertainty; $\sigma_W$ is an additional white noise jitter term; $\sigma_{C}$ is a calibration noise term; $\delta(k,l)$ equals one if measurements $k$ and $l$ are taken within the same night and zero otherwise; $\sigma_{R}$ and $\tau_R$ define a correlated term to model contributions due to granulation \citep{Cegla2019} or instrumental effects, as done in \cite{Hara2020}; $\sigma_{act}, \tau_{act} $, and $P_\mathrm{act}$ are the hyper-parameters of a quasi-periodic covariance term to model the stellar activity.

\begin{figure}
\centering
\includegraphics[width=\linewidth]{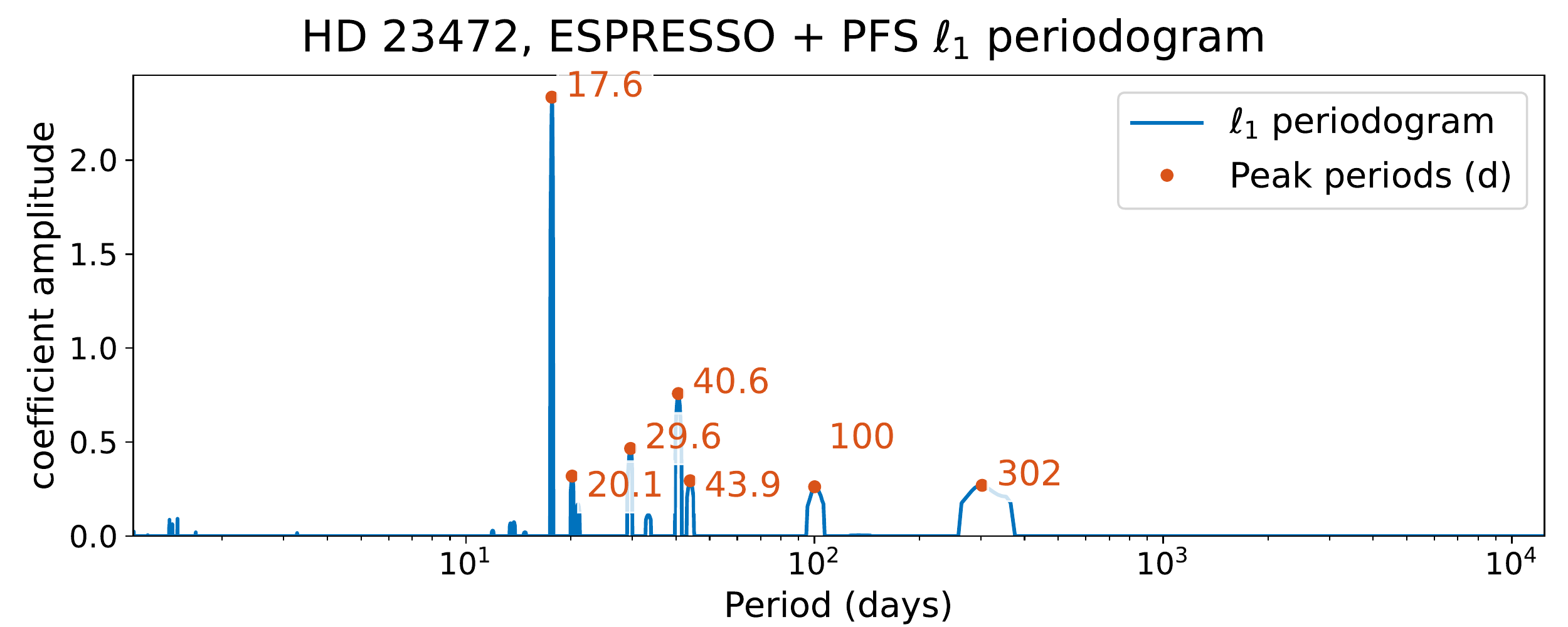}
\caption{$\ell_1$ periodogram with free offsets and covariance models with the highest cross-validation score.}
\label{fig:l1_perio}
\end{figure}

For $\sigma_{W}, \sigma_{R}, \sigma_{QP}$, we use the grid of values $ 0.,0.25, 0.5,0.75,1, 1.25, 1.5,1.75, 2$ m/s, and 0.25, 0.5 m/s for $\sigma_{C}$. $\tau= 0, 3$ or 6 days, $P_{act}$ is fixed to 40 days based on the analysis of activity indicators and  $\tau_{act} = 40, 80, 120$ days, approximately two rotation periods. We try every combination of these values for a total of 7744 models, and find that the model with highest cross-validation has $\sigma_{W} =1$ m/s,  $\sigma_{R} = 0$ m/s, $\sigma_{C} = 0.25$ m/s,   $\sigma_{QP} =0.25$ m/s, and $\tau_{act} = 120$ days. We use a free offset for each dataset (ESPRESSO and PFS).
The corresponding $\ell_1$ periodogram is shown in Figure~\ref{fig:l1_perio}. We find  peaks at 17.64, 40.66, 29.67, 20.16, 43.9, 302 and 100 days with false alarm probabilities of $3.73\cdot10^{-23}$, $1.91\cdot10^{-4}$, $1.09\cdot10^{-3}$, $3.28\cdot10^{-2}$, $2.65\cdot10^{-1}$, $9.65\cdot10^{-2}$, $1.21\cdot10^{-1}$. The signals at 17.64 and 29.67 days correspond to transiting planets. Given that 40 day signals are also present in the ancillary indicators, we interpret the signals at 40.66, 43.9 and 20.16 days as the stellar rotation period and its first harmonic. The 302 and 100 day signals have a FAP of ~10\% and thus cannot be confirmed, but they are viable planetary candidates. The other transiting planets do not show up in the $\ell_1$ periodogram due to their small amplitudes. Some of the noise models with highest cross validation scores also exhibit a peak at 130 days. Interestingly, this signal appears when analysing the ESPRESSO data only and detrending with ancillary indicators.

\end{appendix}

\end{document}